\documentclass[pdflatex,sn-mathphys-num]{sn-jnl}

\usepackage{graphics,epsf}

\usepackage{graphicx}
\usepackage{dcolumn}
\usepackage{bm}
\usepackage{fullpage}
\usepackage{amsmath}
\usepackage{amssymb}
\usepackage{amsfonts}

\usepackage{epstopdf}
\usepackage[T1]{fontenc}
\usepackage[latin9]{inputenc}
\usepackage{amsbsy}
\usepackage{bbm}
\usepackage{braket}
\usepackage{xcolor}
\usepackage{float}
\usepackage[normalem]{ulem}
\usepackage{cancel}

\usepackage{natbib}
\usepackage{graphicx}
\usepackage{color}
\usepackage{gensymb}
\usepackage{float}
\usepackage{amsmath}
\usepackage{tabularx,graphicx}
\usepackage{epstopdf}
\usepackage{latexsym}
\usepackage{color, colortbl}
\usepackage{psfrag}
\usepackage{bbm}
\usepackage{bm}
\usepackage{blindtext}
\usepackage{dsfont}
\usepackage{feynmp}
\usepackage{slashed}
\usepackage{multirow}
\usepackage{appendix}
\usepackage{ragged2e}
\usepackage{bbold}
\usepackage{braket,mleftright}
\usepackage{graphicx}%
\usepackage{multirow}%
\usepackage{amsmath,amssymb,amsfonts}%
\usepackage{amsthm}%
\usepackage{mathrsfs}%
\usepackage[title]{appendix}%
\usepackage{xcolor}%
\usepackage{textcomp}%
\usepackage{manyfoot}%
\usepackage{listings}%
\usepackage{cleveref}

\newcommand{\RVS}{\rm{RbV$_3$Sb$_{5-x}$Sn$_x$}}

\begin{document}

\title{Direct Evidence of Unconventional Superconductivity in Doped Kagome system RV$_3$Sb$_5$}

\author*[1]{\fnm{Avior} \sur{Almoalem}}\email{avior@illinois.edu}

\author[1]{\fnm{Yuqing} \sur{Xing}}

\author[2]{\fnm{Iksu} \sur{Jang}}

\author[2]{\fnm{Daniel J.} \sur{Schultz}}

\author[4]{\fnm{Grgur} \sur{Palle}}

\author[3]{\fnm{Andrea N. Capa} \sur{Salinas}}

\author[3]{\fnm{Stephen D.} \sur{Wilson}}

\author[4]{\fnm{Rafael M.} \sur{Fernandes}}

\author[2]{\fnm{J\"{o}rg} \sur{Schmalian}}
\author*[1]{\fnm{Vidya} \sur{Madhavan}}\email{vm1@illinois.edu}

\affil[1]{\orgdiv{Department of Physics and Materials Research Laboratory, Grainger College of Engineering}, \orgname{University of Illinois at Urbana-Champaign}, \orgaddress{\city{Urbana}, \state{IL}, \country{USA}}}

\affil[2]{\orgdiv{Institute for Theoretical Condensed Matter Physics}, \orgname{Karlsruhe Institute of Technology}, \orgaddress{\city{Karlsruhe}, \country{Germany}}}

\affil[3]{\orgdiv{Materials Department}, \orgname{University of California Santa Barbara}, \orgaddress{\city{Santa Barbara}, \state{CA}, \country{USA}}}

\affil[4]{\orgdiv{Department of Physics and Anthony J. Leggett Institute for Condensed Matter Theory, The Grainger College of Engineering}, \orgname{University of Illinois at Urbana-Champaign}, \orgaddress{\city{Urbana}, \state{IL}, \country{USA}}}

\abstract{The superconducting pairing symmetry of kagome metals remains a central unresolved question largely because phase-sensitive experiments capable of distinguishing between different possible order parameters have been difficult to implement. The spatial and energetic characteristics of impurity bound states measured by spectroscopic-imaging scanning tunneling microscopy encode information on the superconducting order parameter. Here, we use impurity-bound state spectroscopy to investigate optimally doped RbV$_3$Sb$_{5}$ where the charge-density wave instability is fully suppressed. We find that the superconducting state is fully gapped and exhibits two distinct energy scales, consistent with a multiband order parameter. Atomic-scale spectroscopy around nonmagnetic defects reveals pronounced particle-hole asymmetric bound states. Comparison with theoretical calculations demonstrates that these bound states are incompatible with conventional $s$-wave and sign-changing $s^{\pm}$ pairing. Our calculations also show that the data are fully consistent with a chiral order parameter. Our results establish impurity-bound-state spectroscopy as a powerful phase-sensitive probe of superconductivity in kagome materials and provide strong evidence that optimally doped RbV$_3$Sb$_{5}$ realizes a fully gapped chiral superconducting state.}


\maketitle
\newpage
\section{Introduction}\label{sec:introduction}

The kagome superconductors AV$_3$Sb$_5$ (A = Cs, Rb, K) have emerged as a versatile platform for studying intertwined charge-ordered and superconducting phases\cite{wilson2024v3sb5, ortiz2019new, xu2021multiband, xu2022three, jiang2021unconventional, di2026kagome, wang2026unconventional, asaba2024evidence, chen2021roton, li2022rotation, lan2026common, kang2023charge, lin2021complex, le2024superconducting, xing2024optical, wang2023anomalous, yang2020giant, yu2021concurrence, guguchia2023tunable, mielke2022time, khasanov2022time, kenney2021absence, song2022orbital, jiang2023kagome, du2021pressure,Fernandes2026loop}. Although superconductivity is firmly established in this family, the pairing symmetry remains unresolved, partly because different superconducting states may be stabilized across the phase diagram\cite{yang_titanium_2022, chen_highly_2021,wilson2024v3sb5,di2026kagome,schultz_superconductivity_2026}. Resolving the pairing symmetry requires experimental probes that are sensitive to the phase of the superconducting order parameter, a longstanding challenge in the field. 

RbV$_3$Sb$_5$ undergoes a charge-density-wave (CDW) transition at $T_{\mathrm{CDW}}=104$ K, followed by a superconducting transition at $(T_c)_{x=0}=0.9$ K\cite{yin_superconductivity_2021}. Hole doping, achieved in \RVS\ by replacing Sb with Sn, progressively suppresses the CDW, with long-range charge order disappearing at $x=0.1$\cite{oey2022tuning}. Consequently, superconductivity at $x=0.3$ develops from a normal state without CDW order. Hole doping also shifts the Fermi level toward the Van Hove singularity and enhances the superconducting transition temperature to $(T_c)_{x=0.3}=4.2$ K\cite{oey2022tuning}. Although superconductivity in \RVS\ has not been extensively studied in the regime where the CDW is fully suppressed, the closely related hole-doped compound CsV$_{3-x}$Ti$_x$Sb$_5$ exhibits a nodeless and nearly isotropic superconducting gap\cite{yang_titanium_2022, xie_conventional_2024, zhong_nodeless_2023} with clear multigap character\cite{grant2025superconducting, xu2021multiband, gupta2022microscopic}, and an increase in $\mu$SR relaxation rate at $T_c$, usually interpreted as evidence for spontaneous broken time-reversal symmetry (BTRS) in the superconducting state \cite{deng_evidence_2024}. Nuclear magnetic resonance (NMR) measurements, available for CsV$_3$Sb$_5$, reveal that the $^{121}$Sb Knight-shift is suppressed below $T_c$, consistent with spin-singlet pairing\cite{mu2021s}.

Theoretical studies have proposed a range of superconducting pairing symmetries for kagome systems, generally identifying conventional $s$-wave, sign-changing $s$-wave, and chiral $d+id$ as the leading candidates\cite{kiesel_unconventional_2013, wu2021nature, romer_superconductivity_2022, ritz2023superconductivity, li_loop-current_2025, mitra_interplay_2025, alkorta_symmetry-broken_2025,schultz_superconductivity_2026}. Since these are fully gapped pairing states, distinguishing between them requires a probe that is directly sensitive to the phase of the superconducting order parameter. Impurity-induced bound states (Imp-BS) provide precisely such a probe. In fully gapped superconductors they appear as localized in-gap bound states, and are readily accessible by scanning tunneling microscopy (STM)\cite{balatsky_impurity-induced_2006}. The canonical example is bound states created by magnetic impurities embedded in a conventional $s$-wave superconductor\cite{yazdani1997probing,menard2015coherent}, commonly known as Yu-Shiba-Rusinov (YSR) states\cite{yu_luh_bound_1965, shiba_classical_1968, rusinov_superconductivity_1969}.  However, Imp-BS can also arise from nonmagnetic-impurity scattering in unconventional superconductors. Well-known examples are the bound states induced by charge impurities in sign-changing cuprate superconductors like BSCCO\cite{pan2000imaging}, and interband scattering by charge impurities in $s^{\pm}$ superconductors\cite{onari_violation_2009, senga_impurity-induced_2009, tsai_impurity-induced_2009, grothe_bound_2012}. Most relevant to the current study, recent theoretical studies predict that charge impurities in a chiral $d+id$ superconductor generate highly distinctive Imp-BS\cite{wu2026microscopic, cai_deciphering_2025, bunney_chiral_2025, holbaek_interplay_2025}, thus providing a unique fingerprint among the proposed fully gapped pairing states.

\section{Results}\label{sec:results}

\begin{figure}[!ht]
\includegraphics[trim= 0cm 0cm 0cm 0cm,clip=true,width=1\textwidth]{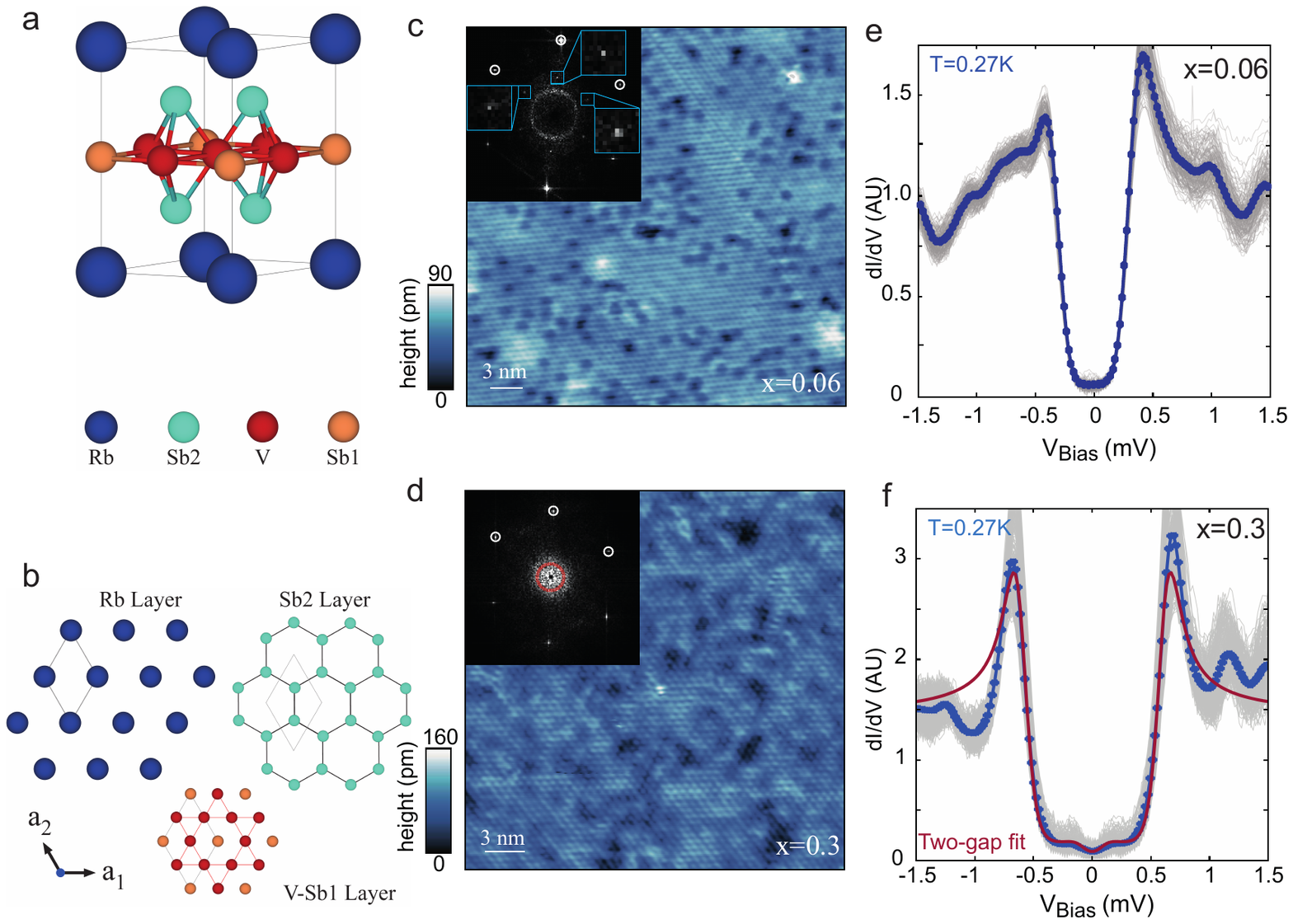}
\caption{\textbf{Crystal structure, topography, and superconducting spectroscopy of Sn-doped RbV$_3$Sb$_{5-x}$Sn$_x$.}
	\textbf{a} Crystal structure of RbV$_3$Sb$_5$. \textbf{b} Top view of the crystal structure showing the triangular Rb layer (blue), Sb honeycomb layer (cyan), and V--Sb kagome layer (red and brown). \textbf{c} STM topography of RbV$_3$Sb$_{4.94}$Sn$_{0.06}$ after removal of surface Rb adatoms, exposing the Sb honeycomb lattice and isolated Sn dopants (35$\times$35~nm$^2$, $V_S=-20$~mV, $I_t=200$~pA). Inset: Fourier transform showing Bragg peaks (white circles), CDW peaks (cyan squares), and Friedel oscillations arising from surface scatterers. \textbf{d} STM topography of RbV$_3$Sb$_{4.7}$Sn$_{0.3}$ (32$\times$32~nm$^2$, $V_S=10$~mV, $I_t=100$~pA). The CDW peaks are absent, while the increased dopant concentration is clearly resolved. Inset: Fourier transform showing the Bragg peaks and the expected Friedel scattering vector associated with the $\Gamma$-centered pocket (red circle). \textbf{e} Spatially averaged tunneling spectrum (blue) together with the individual spectra (gray) acquired on the surface shown in \textbf{c} at $T=0.27$~K. The spectrum is consistent with a predominantly single-gap superconducting state ($V_S=-4$~mV, $I_t=300$~pA, $V_{\mathrm{mod}}=30~\mu$V). \textbf{f} Spatially averaged tunneling spectrum (blue) and individual spectra (gray) measured on the surface shown in \textbf{d}. A two-gap fit (red) reproduces the additional coherence peaks within the superconducting gap, demonstrating the evolution from a predominantly single-gap spectrum to a two-gap superconducting state upon hole doping ($T=0.27$~K, $V_S=-3$~mV, $I_t=150$~pA, $V_{\mathrm{mod}}=30~\mu$V).
}
\label{fig:multiband_SC}
\end{figure}

Single crystals of \RVS\ spanning several doping levels were cleaved \emph{in situ} at 77~K under ultra-high vacuum ($<5\times10^{-10}$ torr) and subsequently transferred to the STM operating at a nominal temperature of $T=270$~mK. The crystal structure consists of a central V-Sb kagome layer sandwiched between two Sb honeycomb layers and separated by triangular Rb layers, as illustrated in Fig.~\ref{fig:multiband_SC}a,b. Cleavage always occurs between the Rb and Sb layers, with the STM measurements performed on the exposed Sb honeycomb layer (Fig.~\ref{fig:multiband_SC}a-d) (see Supplementary Fig.~1).

Although the V-Sb kagome layer is not directly exposed after cleavage, its electronic states, including superconductivity, are readily accessed through the terminating Sb honeycomb layer\cite{jiang2021unconventional,chen2021roton,xing2024optical,di2026kagome,xu2021multiband,huang2025spatially,huai2025electronic,xu2025pervasive,zhao2021cascade,tu2025symmetry,li2023unidirectional,li2022rotation,wang2021charge,nie2022charge}. A defining characteristic of the optimally doped ($x=0.3$) surface is the complete absence of charge-density-wave (CDW) order, as demonstrated by comparing its topography and corresponding Fast Fourier transform (FFT), with those of the moderately doped sample $x=0.06$, as shown in Fig.~\ref{fig:multiband_SC}c,d. This observation is consistent with transport measurements showing that long-range CDW order is fully suppressed at this doping level\cite{oey2022tuning}. We also measure an increase in the superconducting transition temperature from $T_c=0.7$~K in the parent compound to $T_c=3.7$~K at $x=0.3$, close to the reported value of 4.2~K\cite{oey2022tuning}. Simultaneously, the upper critical field along the $c$ axis increases from $H_{c2}=0.02$~T to $2$~T (see Supplementary Fig.~2 for field and temperature dependence of $x=0.3$). Finally, we find that Sn atoms predominantly substitute for Sb atoms in the V-Sb kagome plane, as seen through topography of isolated dopant sites (Supplementary Fig.~3) and Friedel oscillations data (Supplementary Fig.~4). This conclusion is further supported by recent nuclear quadrupole resonance measurements on doped CsV$_3$Sb$_{5-x}$Sn$_x$\cite{k2026observation,oey2022fermi}.

Examining the evolution of the superconducting spectra reveals a stark change in the local density of states (LDOS) signal. For $x=0.06$, well-defined coherence peaks appear at $\pm420~\mu$eV. Individual spectra shown in Fig.~\ref{fig:multiband_SC}e, and acquired at 100 locations across the topography shown in Fig.~\ref{fig:multiband_SC}c, show that the gap remains unchanged on individual Sn dopant sites. The averaged spectrum shows a fully gapped superconducting state, apart from a small residual density of states ($\sim10\%$) at the Fermi level. Although this spectrum is well described by a single-gap superconducting state, the finite zero-bias conductance prevents us from excluding a second unresolved gap. Such residual spectral weight may originate from a second gap that is only partially resolved because of experimental resolution, or from pair-breaking processes\cite{moreno2025gapless,4hbthermal}. The possibility of multigap superconductivity is consistent with several previous studies\cite{hossain2025unconventional,wilson2024v3sb5,yin2021strain,gupta2022microscopic}, although single-gap behavior has also been reported\cite{xu2021multiband,di2026kagome,chen2021roton,liang2021three}. 

The optimally doped compound exhibits qualitatively different behavior. Pronounced coherence peaks at $\pm680~\mu$eV are accompanied by additional shoulder-like coherence features at $\pm180~\mu$eV, indicating the presence of a second superconducting energy scale. Low-temperature measurements of the parent compound (Supplementary Fig.~5), and a Pb crystal (Supplementary Fig.~6), demonstrate our ability to resolve small gaps. Importantly, the secondary coherence peaks persist throughout the entire spectroscopic line cut shown in Fig.~\ref{fig:multiband_SC}f and Supplementary Fig.~7, demonstrating that they are an intrinsic property of the superconducting state. Throughout this work, we therefore describe the spectra using a two-gap model with $\Delta_1=680~\mu$eV and $\Delta_2=180~\mu$eV. 

The two superconducting gaps naturally arise from the multiband electronic structure of AV$_3$Sb$_5$. Independent evidence for two superconducting gaps is also obtained from quasiparticle interference (QPI) measurements, which provide momentum-space confirmation of the two-gap interpretation. QPI analysis shows that the small planar Sb-derived Fermi-surface pocket centered at $\Gamma$ acquires the larger gap, $\Delta_1=680~\mu$eV, consistent with previous QPI measurements on RbV$_3$Sb$_5$\cite{yan_chiral_2024}. This assignment is further supported by pressure studies, where suppression of superconductivity accompanies a Lifshitz transition that removes the $\Gamma$-centered pocket\cite{bhandari_first-principles_2024, tsirlin_effect_2023, si_charge_2022, du2021pressure, zhang_pressure-induced_2021}. The smaller gap, $\Delta_2=180~\mu$eV, is therefore associated primarily with the larger V-derived Fermi-surface sheets (together with contributions from apical Sb orbitals), which continue to produce a QPI signal over the energy range $\Delta_2<V_{\mathrm{bias}}<\Delta_1$ (Supplementary Fig.~8).

\subsection{Type-I impurity-induced bound states}\label{sec:bound_state_1}

\begin{figure}[!ht]
\includegraphics[trim= 0cm 0cm 0cm 0cm,clip=true,width=1\textwidth]{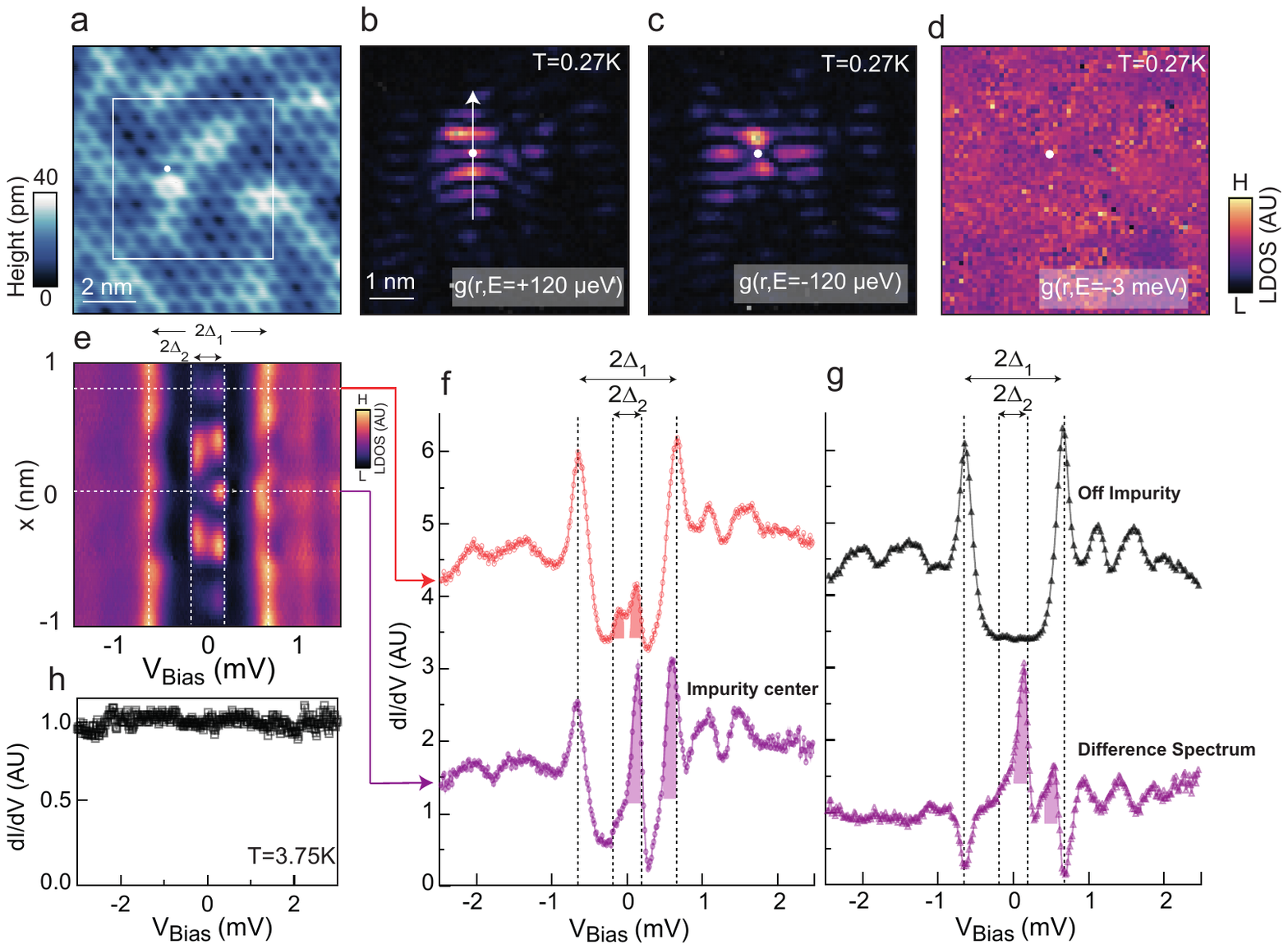}
\caption{\textbf{Spatial and spectroscopic characteristics of type-I Imp-BS.}
	\textbf{a} STM topography of the Sb honeycomb lattice in the vicinity of a type-I state. No structural defects are resolved at the bound-state center (white dot), indicating that the impurity resides beneath the terminating Sb layer. The white square marks the region shown in panels \textbf{b-d} ($V_S=-3$~V, $I_t=5$~nA). \textbf{b,c} Differential conductance maps acquired at $\pm120~\mu$eV, corresponding to the dominant impurity-induced states. Pronounced spatial oscillations and a clear reduction of the native sixfold rotational symmetry are observed. The white arrow indicates the line cut shown in panel \textbf{e} ($V_S=-3$~mV, $I_t=150$~pA, $V_{\mathrm{mod}}=30~\mu$V). \textbf{d} Differential conductance map acquired at $-3$~meV, outside the superconducting gap. No anomaly is observed at the impurity position, indicating that the bound state is confined to the superconducting state rather than arising from normal-state electronic inhomogeneity ($V_S=-3$~mV, $I_t=150$~pA, $V_{\mathrm{mod}}=30~\mu$V). \textbf{e} Spatial evolution of the tunneling spectra along the line indicated in panel \textbf{b}. Dashed lines mark the superconducting gap energies $\Delta_1=680~\mu$eV and $\Delta_2=180~\mu$eV. \textbf{f} Tunneling spectra acquired at the impurity center (purple) and away from it (orange), corresponding to the horizontal dashed lines in panel \textbf{e}. The impurity suppresses the superconducting coherence peaks and produces two Imp-BS. The particle-hole asymmetry is strongest at the impurity center and gradually evolves toward a more symmetric spectral weight distribution with increasing distance. Black dashed lines indicate the superconducting gap energies. \textbf{g} Background-subtracted spectrum (purple) obtained by subtracting the bare superconducting spectrum (black), measured away from the impurity, from the spectrum acquired at the impurity center. The subtraction highlights the states at $E_{\mathrm{Imp-BS}}^{(2)}=135~\mu$eV and $E_{\mathrm{Imp-BS}}^{(1)}=520~\mu$eV. The pronounced particle-hole asymmetry of the impurity-induced states is consistent with the behavior expected for a nonmagnetic impurity in a chiral $d+id$ superconductor (see main text). \textbf{h} Tunneling spectrum acquired at the impurity site above the superconducting transition temperature. The absence of Kondo-like features, together with the nearly symmetric normal-state density of states, indicates that the impurity is nonmagnetic. This contrasts with the Kondo resonances previously reported for magnetic impurities in CsV$_3$Sb$_5$. Panels \textbf{b-h}: $V_S=-3$~mV, $I_t=150$~pA, $V_{\mathrm{mod}}=30~\mu$V.
}
\label{fig:defect_type1}
\end{figure}

Having established the multiband character of the superconducting state, we next investigate Imp-BS that emerge in the superconducting state. We identify two distinct types of Imp-BS. As we show below, the corresponding spectroscopic and spatial behaviors, together with the existence of both types of in-gap states, are inconsistent with a conventional $s$-wave order parameter.

We first characterize the atomic structure surrounding a representative type-I Imp-BS. Figure~\ref{fig:defect_type1}a shows a high-bias STM topography in the vicinity of the impurity. Imaging at high bias minimizes contrast arising from the energy-dependent LDOS associated with nearby dopant atoms, allowing the atomic structure to be resolved more clearly. The honeycomb lattice is intact, with no missing atoms or other structural irregularities at the impurity position. These observations suggest that the impurity is most likely located within the underlying V-Sb kagome layer, although we cannot exclude the possibility that the scattering potential originates from a small fraction of Sn atoms substituting for Sb in the honeycomb layer.

The electronic signature of the impurity is revealed only inside the superconducting gap. Differential conductance maps acquired at $\pm120~\mu$eV are shown in Figs.~\ref{fig:defect_type1}b,c. At these energies, the impurity gives rise to pronounced in-gap spectral weight accompanied by clear spatial oscillations, a strong reduction of the native sixfold rotational symmetry to an approximately twofold pattern, and markedly different particle- and hole-like intensity distributions. In contrast, a conductance map acquired well outside the superconducting gap ($-3$~meV) exhibits no detectable anomaly (Fig.~\ref{fig:defect_type1}d), demonstrating that the observed contrast is intrinsic to the superconducting state rather than arising from normal-state electronic inhomogeneity. 

To examine the spatial evolution of the Imp-BS, we acquired tunneling spectra along the direction indicated by the arrow in Fig.~\ref{fig:defect_type1}b. The resulting spectra reveal a striking spatial evolution of the Imp-BS (Fig.~\ref{fig:defect_type1}e). At the impurity center ($x=0$ in the figure), the Imp-BS exhibits pronounced particle-hole asymmetry. As the distance from the impurity increases, the spectral weight gradually redistributes between the particle- and hole-like branches, recovering an increasingly particle-hole symmetric line shape. Representative spectra acquired at the impurity center and far from the defect are compared in Fig.~\ref{fig:defect_type1}f.

The tunneling spectrum measured directly above the impurity (purple spectrum in Fig.~\ref{fig:defect_type1}f) contains two impurity-induced states. The first appears at $E_{\mathrm{Imp-BS}}^{(2)}=+135~\mu$eV, well inside the smaller superconducting gap ($E_{\mathrm{Imp-BS}}^{(2)}<\Delta_2=180~\mu$eV), while the second is located at $E_{\mathrm{Imp-BS}}^{(1)}=+585~\mu$eV, between the two superconducting gaps ($\Delta_2<E_{\mathrm{Imp-BS}}^{(1)}<\Delta_1=680~\mu$eV). Both states lie close to their corresponding superconducting coherence peaks. Simultaneously, the superconducting coherence peaks are substantially suppressed at the impurity site, indicating a transfer of spectral weight from the superconducting condensate into the Imp-BS. Another important feature is the maximal particle-hole asymmetry at the impurity center, which is progressively restored to particle-hole symmetry with increasing distance from the defect. 

To isolate the impurity contribution to the tunneling spectrum, we subtract the background tunneling spectrum acquired far from the defect. As shown in Fig.~\ref{fig:defect_type1}g, this procedure enhances the visibility of both states and shifts the apparent energy of the higher-energy state from $E_{\mathrm{Imp-BS}}^{(1)}=585~\mu$eV to $E_{\mathrm{Imp-BS}}^{(1)}=520~\mu$eV. The lower-energy resonance at $E_{\mathrm{Imp-BS}}^{(2)}$ remains visible even in zero-bias conductance maps (Supplementary Fig.~9), highlighting its existence at $E<\Delta_2$.

Interpreting these bound states requires identifying the nature of the underlying impurity potential. We therefore characterize the defects in the normal state to determine whether they are magnetic or nonmagnetic. First, the density of impurities which generate Imp-BS is substantially lower than the Sn dopant concentration, indicating that these impurities are not associated with intentional Sn substitution within the kagome layer. Second, although there is no reason to expect the impurity to be magnetic because AV$_3$Sb$_5$ compounds lack local spins\cite{wilson2024v3sb5,di2026kagome}, we further test this possibility by searching for the signature of a Kondo resonance in the vicinity of the impurity. Indeed, strong Kondo resonances have been observed for deliberately introduced magnetic impurities in kagome materials \cite{huang2025spatially,tu2025symmetry,zheng2025quasiparticle}. We therefore show in Figure~\ref{fig:defect_type1}h the tunneling spectrum acquired at the impurity in the normal state. The spectrum is essentially featureless and exhibits neither Kondo-like resonances nor appreciable particle-hole asymmetry, consistent with the conductance map acquired outside the gap. This is in clear contrast to the characteristic Fano line shape associated with magnetic impurities in kagome superconductors\cite{huang2025spatially,tu2025symmetry,zheng2025quasiparticle}, supporting a nonmagnetic origin of the impurity. Although magnetic impurities without observable Kondo resonances have been reported in other material systems, such as Fe impurities in NbSe$_2$\cite{menard2015coherent}, Kondo signatures from magnetic impurities are well established in the Kagome family of materials\cite{huang2025spatially,tu2025symmetry,zheng2025quasiparticle}. We also note that magnetic impurities originating from magnetic ions contaminating the sample, as in NbSe$_2$\cite{menard2015coherent}, are unlikely, because we would expect YSR states to appear in samples at lower doping, where they are in fact absent.

\subsection{Theoretical analysis of Imp-BS} \label{sec:theory_modelling}

\begin{figure}[!ht]
	\centering
	\includegraphics[trim=0cm 0cm 0cm 0cm,clip=true,width=1\textwidth]{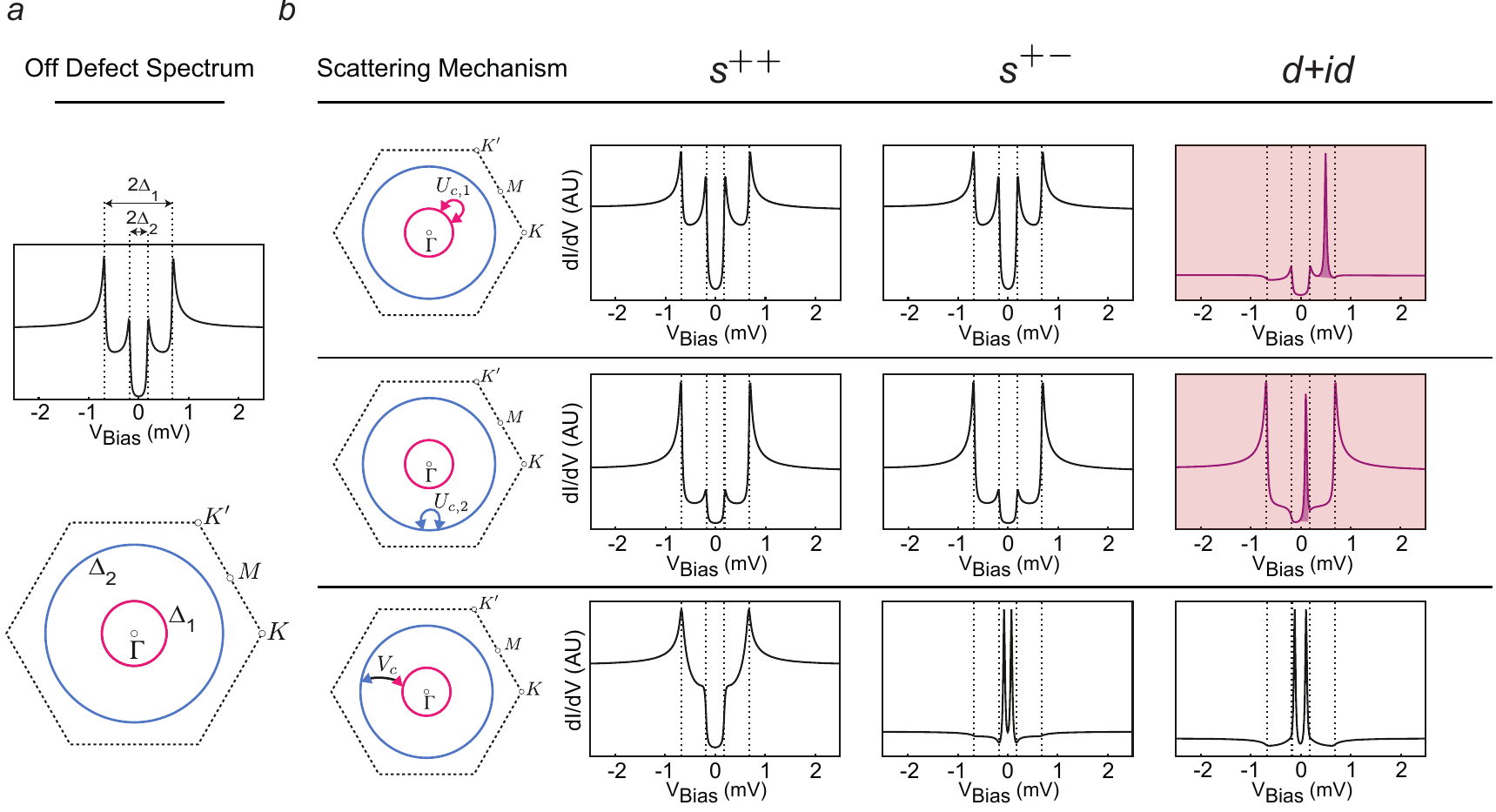}
	\caption{\textbf{Theoretical response of candidate superconducting pairing symmetries to a nonmagnetic impurity.}
		\textbf{a} Density of states obtained from the model calculations without impurity, together with the corresponding multiband Fermi surface. The larger superconducting gap, $\Delta_1$, is assigned to the Sb-derived $\Gamma$ pocket, whereas the smaller gap, $\Delta_2$, resides on the V-derived Fermi-surface sheets, consistent with the experimental observations. \textbf{b} Calculated local density of states at the impurity center for the candidate pairing symmetries and nonmagnetic scattering channels indicated in the left column. The highlighted panels correspond to the only pairing states that reproduce the key experimental observations within the present theoretical framework, namely the pronounced particle-hole asymmetry and redistribution of spectral weight observed for the type-I impurity-induced state. Both correspond to a chiral $d+id$ superconducting state with different intraband scattering potentials. The numerical parameters used in the calculations, together with an extended comparison including additional nonmagnetic- and magnetic-scattering channels, are provided in the Supplementary Information.}
	\label{fig:table}
\end{figure}

To identify the superconducting pairing symmetry underlying the observed Imp-BS, we perform microscopic calculations for the three candidate pairing symmetries that give fully gapped spectra: conventional $s$-wave (denoted $s^{++}$), sign-changing $s$-wave (denoted $s^{\pm}$), and chiral $d+id$. Our focus on spin-singlet pairing states is based on a previous NMR study of CsV$_3$Sb$_5$ that is consistent with spin-singlet superconductivity\cite{mu2021s}.

According to the two-gap superconducting spectra, we describe the electronic structure using a minimal two-band Hamiltonian,

\begin{align}
	H ={}& \sum_{\bm{k},a}\left[\xi_{\bm{k},a} c^\dagger_{\bm{k}\sigma a} c_{\bm{k}\sigma a} + \left(\Delta_a(\bm{k}) c^\dagger_{\bm{k}\uparrow,a} c^\dagger_{-\bm{k},\downarrow,a} +\mathrm{H.c.}\right)\right],
\end{align}
where $\xi_{\bm{k},a}$ denotes the dispersion of band $a$ ($a=1,2$). The two bands, shown in Fig.~\ref{fig:table}a, model the Sb-dominated Fermi pocket ($a=1$) and the V-dominated Fermi surface ($a=2$). 
The gap functions of the three candidate pairing states considered here are given by:

\begin{align}
	&s^{++}: \begin{cases}
		\Delta_1(\bm{k}) = |\Delta_1| \\
		\Delta_2(\bm{k}) = |\Delta_2|
	\end{cases},\quad
	s^{\pm}:\begin{cases}
		\Delta_1(\bm{k}) = +|\Delta_1| \\
		\Delta_2(\bm{k}) = -|\Delta_2|
	\end{cases}
	\\
	&d+id:\begin{cases}
		\Delta_1(\bm{k}) = |\Delta_1|\left[\cos k_x-\cos k_y+i\sin k_x\sin k_y\right] \\
		\Delta_2(\bm{k}) = |\Delta_2|\left[\cos k_x-\cos k_y+i\sin k_x\sin k_y\right].
	\end{cases}
\end{align}

We set the gap values $\Delta_1$ and $\Delta_2$  to agree with those observed experimentally. As a result, the LDOS has the same form for all three pairing states, shown in Fig.~\ref{fig:table}a. The impurity is modeled through a local scattering potential, which generally includes both nonmagnetic (charge) and magnetic components. As shown in the first column of Fig.~\ref{fig:table}b, the potential contains terms that scatter states within the same band (intraband potentials $U_{c,1}$ and  $U_{c,2}$) and between the two bands (interband potential $V_{c}$). For each pairing symmetry, we calculate the LDOS at the impurity site for a broad range of scattering strengths (see Methods for more details). Complete results are presented in Supplementary Information, while Fig.~\ref{fig:table}b highlights the cases relevant to the present experiment, focusing on a nonmagnetic impurity.

The calculated LDOS allows us to immediately distinguish the three candidate pairings. As expected from Anderson's theorem\cite{anderson1959theory}, a nonmagnetic impurity does not generate an Imp-BS in a conventional $s^{++}$ superconductor, as seen in the second column of Fig.~\ref{fig:table}b. For a sign-changing $s^{\pm}$ state, interband nonmagnetic scattering is known to produce an Imp-BS\cite{bang2009impurity,beaird2012impurity}. However, these states remain confined to energies below the small superconducting gap and exhibit particle-hole symmetric spectral weight. Both predictions are inconsistent with experimental observations, which show pronounced particle-hole asymmetry and a second resonance satisfying $\Delta_2<E_{\mathrm{Imp-BS}}<\Delta_1$.

The remaining candidate is the chiral $d+id$ state. In this case, the spectral weight of one member of the particle-hole pair of Imp-BS at the impurity site is proportional to\cite{cai_deciphering_2025,wu2026microscopic}

\begin{align}
	Z_{-}
	=
	\left|
	\int_{\mathrm{BZ}}
	\frac{\Delta_a(\boldsymbol{k})}
	{-\omega^2+|\xi_a(\boldsymbol{k})|^2+|\Delta_a(\boldsymbol{k})|^2}
	\frac{d^2\boldsymbol{k}}{(2\pi)^2}
	\right|^2.\label{eq:Zminums}
\end{align}

For superconducting order parameters whose angular average vanishes and $|\Delta_a(\bm{k})|^2$ is $C_{6z}$-symmetric, this integral is identically zero. Consequently, for $d+id$, one member of the particle-hole pair ($Z_{-}$) is completely suppressed at the impurity center, while the counterpart $Z_{+}$ is not, producing the pronounced particle-hole asymmetry observed experimentally (see Methods for details). The missing spectral weight gradually reappears away from the impurity, naturally explaining the spatial evolution of the tunneling spectra shown in Fig.~\ref{fig:defect_type1}e,f. Importantly, this behavior is independent of the detailed form of the nonmagnetic impurity potential. Taken together, the theoretical calculations identify chiral $d+id$ superconductivity as the leading pairing symmetry consistent with all experimental observations.

\begin{figure}[!ht]
	\includegraphics[trim= 0cm 0cm 0cm 0cm,clip=true,width=1\textwidth]{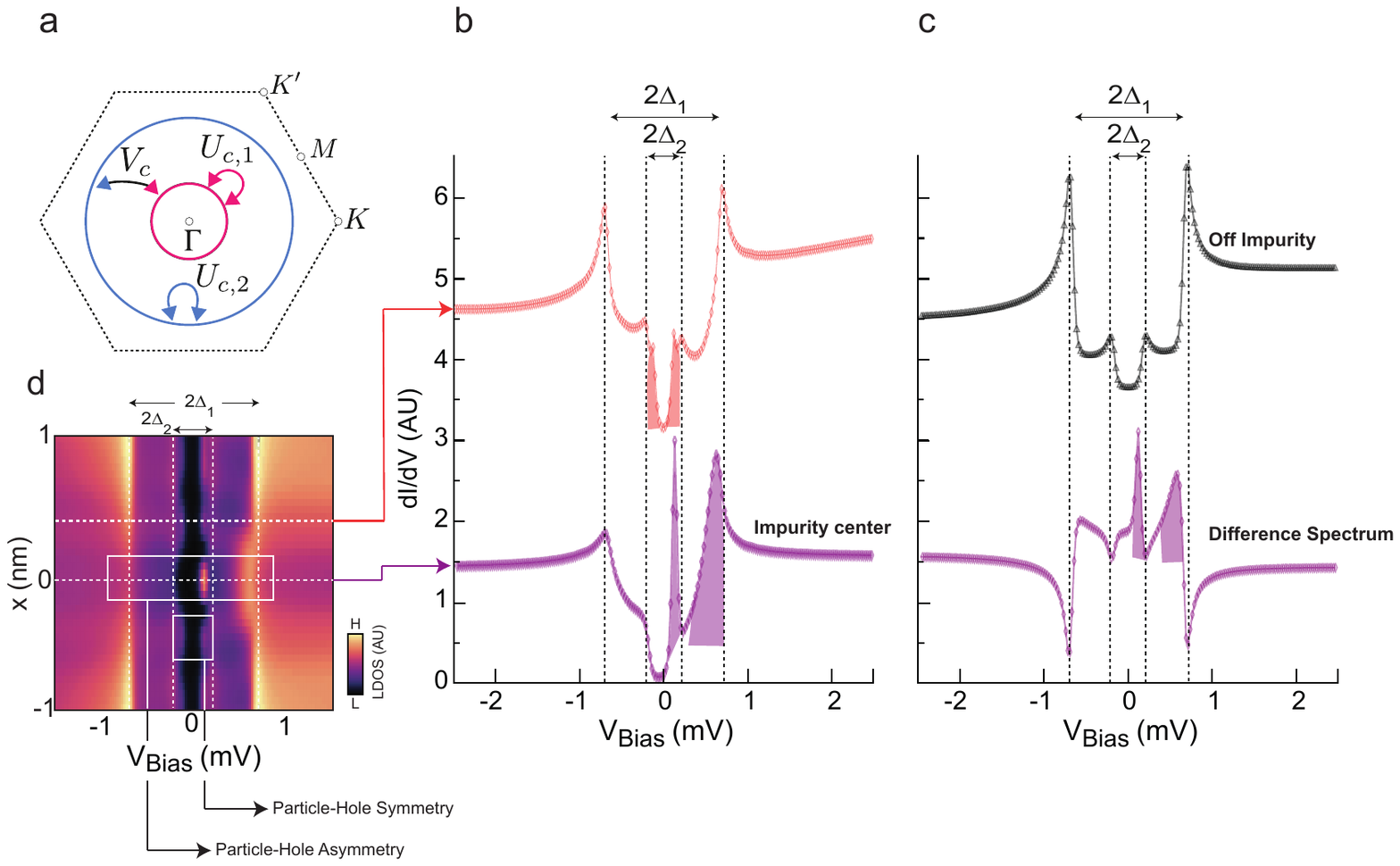}
\caption{\textbf{Microscopic calculation of the type-I impurity-induced state using a nonmagnetic scattering potential.}
	\textbf{a} Schematic of the impurity-scattering processes included in the calculation. Both intraband and interband charge scattering are incorporated. \textbf{b} Calculated local density of states at the impurity center (purple) and away from it (orange), corresponding to the horizontal dashed lines in panel \textbf{d}. The calculation reproduces the pronounced particle-hole asymmetry at the impurity center and the gradual recovery of particle-hole symmetry with increasing distance observed experimentally. Black dashed lines indicate the superconducting gap energies $\Delta_1$ and $\Delta_2$. \textbf{c} Pristine local density of states used in the calculation (black) together with the background-subtracted local density of states at the impurity center (purple). The calculated spectra reproduce the experimental tunneling spectra shown in Fig.~\ref{fig:defect_type1}g, including the energies and relative spectral weights of the impurity-induced resonances. \textbf{d} Calculated local density of states as a function of energy and distance from the impurity center. The simulation reproduces the strong particle-hole asymmetry at $x=0$, the redistribution of spectral weight with increasing distance, the high-energy state adjacent to the $\Delta_1$ coherence peak, and the suppression of the superconducting coherence peaks observed experimentally (Fig.~\ref{fig:defect_type1}e,f). Numerical parameters are given in the Methods.
}
	\label{fig:simulation}
\end{figure}

To go beyond this qualitative analysis, we compute LDOS for the $d+id$ case including both intraband and interband nonmagnetic scattering processes at different locations with respect to the impurity center. As shown in Fig.~\ref{fig:simulation}, the calculated LDOS reproduces the key experimental observations. In particular, the impurity center exhibits maximal particle-hole asymmetry, while the spectra evolve continuously toward a more particle-hole symmetric line shape with increasing distance from the impurity (Fig.~\ref{fig:simulation}b). The background-subtracted local density of states at the impurity center (Fig.~\ref{fig:simulation}c) closely reproduces the experimental tunneling spectrum shown in Fig.~\ref{fig:defect_type1}g, including the resonance energies and their relative spectral weights.

The calculated spatial evolution of the Imp-BS is shown in Fig.~\ref{fig:simulation}d. At the impurity center, both states, $E_{\mathrm{Imp-BS}}^{(2)}<\Delta_2$ and $\Delta_2<E_{\mathrm{Imp-BS}}^{(1)}<\Delta_1$, appear predominantly above the Fermi level. Away from the impurity, the spectral weight of $E_{\mathrm{Imp-BS}}^{(2)}$ oscillates with the Fermi momentum of the corresponding Fermi surface as described in the Methods section. By contrast, the higher-energy state appears as a broad feature at the impurity center, and disappears in the line cut due to the much smaller Fermi momentum of the corresponding Fermi surface. This distinction further supports our assignment of the gaps to the different Fermi surfaces. Particle-hole symmetry of $E_{\mathrm{Imp-BS}}^{(2)}<\Delta_2$ is progressively restored with increasing distance from the impurity, in close agreement with the experimental observations.

\subsection{Type-II impurity-induced bound states}

\begin{figure}[!ht]
\includegraphics[trim= 0cm 0cm 0cm 0cm,clip=true,width=1\textwidth]{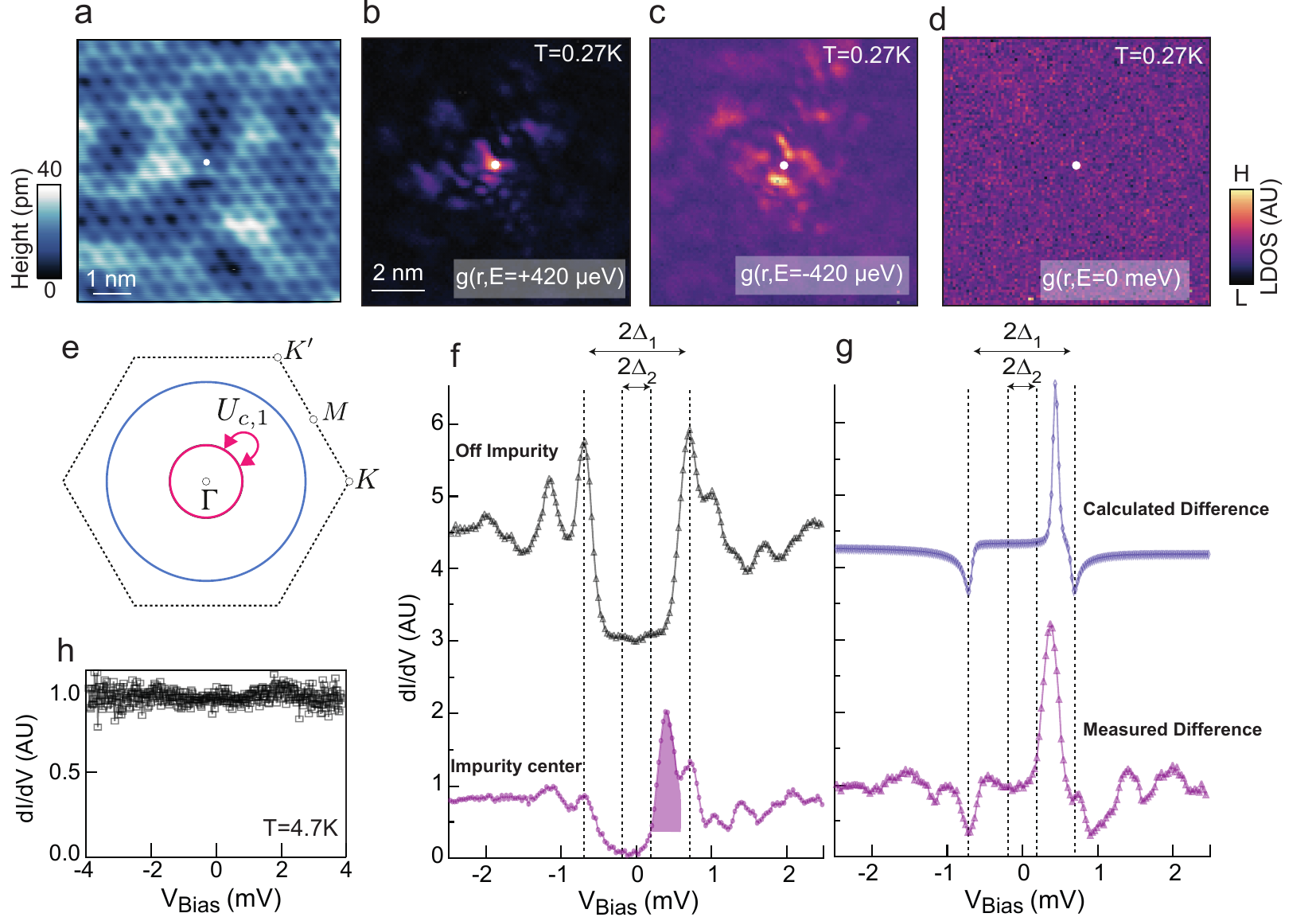}
\caption{\textbf{Spatial and spectroscopic characteristics of the type-II Imp-BS.}
	\textbf{a} STM topography of the Sb honeycomb lattice in the vicinity of a type-II state. No structural defects are resolved at the impurity center (white dot), indicating that the defect resides beneath the terminating Sb layer ($V_S=-3$~V, $I_t=5$~nA). \textbf{b-d} Differential conductance maps acquired at $\pm420~\mu$eV, corresponding to the Imp-BS, and at the Fermi level. The spatial scale bar is identical for panels \textbf{b-d}, which were acquired simultaneously. Unlike the type-I state, the type-II state exhibits no pronounced spatial oscillations, although the maximal particle-hole asymmetry remains localized at the impurity center. Furthermore, no anomaly is observed at the Fermi level ($V_S=-3$~mV, $I_t=150$~pA, $V_{\mathrm{mod}}=30~\mu$V). \textbf{e} Schematic of the impurity-scattering processes used to model the type-II impurity. \textbf{f} Tunneling spectra acquired at the impurity center (purple) and away from it (black). A single bound state appears between the two superconducting gaps ($\Delta_2<E_{\mathrm{Imp-BS}}<\Delta_1$), with no additional states inside the smaller gap. Its energy differs from that of the higher-energy resonance associated with the type-I state, indicating a different impurity-scattering potential. Like the type-I case, the superconducting coherence peaks are suppressed at the impurity site and the spectra exhibit pronounced particle-hole asymmetry. Black dashed lines indicate the superconducting gap energies. \textbf{g} Background-subtracted experimental (bottom) and calculated (top) tunneling spectra obtained by subtracting the pristine superconducting spectrum measured away from the impurity. Subtraction highlights the state at $E_{\mathrm{Imp-BS}}=450~\mu$eV. The calculated spectrum reproduces the pronounced particle-hole asymmetry observed experimentally, consistent with the response of a nonmagnetic charge impurity embedded in a chiral $d+id$ superconductor. \textbf{h} Tunneling spectrum acquired at the impurity site, above the superconducting transition temperature. The absence of Kondo-like features, together with the nearly symmetric normal-state spectrum, indicates that the impurity is nonmagnetic. Panels \textbf{b-f}: $V_S=-3$~mV, $I_t=150$~pA, $V_{\mathrm{mod}}=30~\mu$V. Panel \textbf{h}: $V_S=-4$~mV, $I_t=200$~pA, $V_{\mathrm{mod}}=30~\mu$V.
}
\label{fig:defect_type2}
\end{figure}

For intraband scattering on the Fermi surface hosting the larger gap $\Delta_1$, theoretical analysis predicts the existence of a single Imp-BS state at an energy between the two superconducting gaps, $\Delta_2<E_{\mathrm{Imp-BS}}<\Delta_1$, while retaining the pronounced particle-hole asymmetry. Although such bound states are considerably less common, we identify precisely this behavior experimentally. The observation of such a bound state provides an additional constraint on the superconducting pairing symmetry. The existence of this type-II bound state is completely incompatible with an $s^{\pm}$ state. This is because, as shown in Fig.~\ref{fig:table}b, impurity-bound states in the $s^\pm$ case are restricted to energies below the smaller superconducting gap ($E_{\mathrm{Imp-BS}}<\Delta_2$) due to the sign change of the gap occurring only between the Fermi surfaces.

As discussed in Sec.~\ref{sec:bound_state_1}, the absence of any structural defect in the high-bias topography (Fig.~\ref{fig:defect_type2}a) indicates that the impurity resides within the underlying V-Sb kagome layer. However, the spatial structure of the bound state differs markedly from that of type-I. Differential conductance maps acquired at the resonance energy ($E_{\mathrm{Imp-BS}}\approx\pm420~\mu$eV) are shown in Figs.~\ref{fig:defect_type2}b,c. Unlike type-I Imp-BS, which exhibits pronounced spatial oscillations and an approximately twofold symmetry, this Imp-BS remains strongly localized and shows no comparable oscillatory behavior.

The conductance map acquired at the Fermi level (Fig.~\ref{fig:defect_type2}d) reveals another distinction. In contrast to type-I, no detectable zero-bias anomaly is observed, indicating that this bound state does not generate spectral weight at the Fermi level. Within our theoretical framework, this behavior is naturally reproduced by a scattering potential dominated by intraband scattering within the smaller Fermi-surface pocket, as illustrated schematically in Fig.~\ref{fig:defect_type2}e.

The tunneling spectrum acquired directly above the impurity shows a single Imp-BS at $E_{\mathrm{Imp-BS}}=450~\mu$eV, satisfying $\Delta_2<E_{\mathrm{Imp-BS}}<\Delta_1$,  Fig.~\ref{fig:defect_type2}f. The calculated background-subtracted spectrum for the $d+id$ state reproduces both the energy and the pronounced particle-hole asymmetry of the experimental data (Fig.~\ref{fig:defect_type2}g). As for the type-I impurity, the featureless normal-state spectrum supports a nonmagnetic origin, with the distinction between the two impurity types emerging only in the superconducting state (Fig.~\ref{fig:defect_type2}h).


\section*{Discussion}

A defining property of conventional $s$-wave superconductors is their robustness against nonmagnetic disorder, known as Anderson's theorem\cite{anderson1959theory}. Because the phase of a conventional s-wave order parameter is uniform over the Fermi surface, nonmagnetic scattering does not generate in-gap bound states. In contrast, unconventional gap functions can average to zero over the Fermi surface, allowing nonmagnetic impurities to act as effective pair breakers that generate Imp-BS\cite{abrikosov1961zh,abrikosov1961zhetf,holbaek2023unconventional}. Therefore, our observations impose stringent constraints on the pairing symmetry of RbV$_3$Sb$_{4.7}$Sn$_{0.3}$. First, STM spectroscopy reveals a fully gapped superconducting state with two distinct energy scales. Second, impurity spectroscopy reveals pronounced particle-hole asymmetry that persists across distinct impurity-scattering potentials. The existence of Imp-BS satisfying $\Delta_2<E_{\mathrm{Imp-BS}}<\Delta_1$ further requires the order parameter to average to zero over the Fermi surface hosting the larger gap, rendering a sign-changing $s$-wave pairing implausible. These observations, together with the microscopic calculations and the assumption of spin-singlet pairing based on an NMR study of CsV$_3$Sb$_5$\cite{mu2021s}, strongly support chiral $d+id$ superconductivity in RbV$_3$Sb$_{4.7}$Sn$_{0.3}$. 

To further substantiate the unconventional nature of the superconducting state, we consider whether purely magnetic impurities could account for our experimental observations. Although asymmetric YSR resonances are known to exist, as observed for Fe substitution in NbSe$_2$\cite{menard2015coherent} or Mn adatoms on Nb\cite{yazdani1997probing}, this behavior does not arise generically. 

One possible origin of particle-hole asymmetry in YSR states is an impurity potential containing both charge and magnetic components. Calculations for selected Fe impurities in NbSe$_2$\cite{menard2015coherent} demonstrate that strongly asymmetric YSR resonances may emerge for a specific tuning of the charge and magnetic scattering strengths. In contrast, all impurities investigated in the present study, including the two distinct impurity types presented in the main text and the three additional examples shown in Supplementary Fig.~12, exhibit pronounced particle-hole asymmetry despite their otherwise very different spatial structures and spectroscopic signatures. 

Asymmetric YSR spectra may also arise from interference between tunneling pathways through a Kondo impurity\cite{madhavan1998tunneling}. Such interference generally produces an asymmetric Fano line shape in the local density of states that remains visible above the superconducting transition temperature. In contrast, neither impurity type investigated here exhibits Kondo-like signatures or appreciable asymmetry in the normal state, which we discuss further in Supplementary Fig.~13. The absence of a detectable Kondo signal, together with the robustness of this asymmetry across multiple impurity potentials, strongly suggests that the observed particle-hole asymmetry is an intrinsic consequence of unconventional superconductivity rather than a manifestation of conventional magnetic-impurity physics.

Charge impurities may also generate Imp-BS in sign-changing $s^{\pm}$ and $s+is$ superconducting states. The experimental observations presented here are incompatible with both scenarios for two separate reasons. First, the pronounced particle-hole asymmetry observed for both impurity classes is not reproduced by generic nonmagnetic or magnetic impurities embedded in an $s^{\pm}$ superconductor. As shown in Supplementary Figs.~10 and 11, both nonmagnetic and magnetic impurities generate two particle-hole symmetric Imp-BS. Second, in this work we observe resonances located between the two superconducting gaps ($\Delta_2<E_{\mathrm{Imp-BS}}<\Delta_1$) (Figs.~\ref{fig:defect_type1}, \ref{fig:defect_type2}). For an $s^{\pm}$ superconductor, however, the standard YSR or nonmagnetic Imp-BS energies are expected to lie below the smaller superconducting gap, since the sign change occurs exclusively between the two Fermi surfaces. The same considerations apply to $s+is$, for which an additional constraint arises. Due to the two $s$-wave components belonging to the same irreducible representation, an $s+is$ state is generally expected to emerge through two successive superconducting phase transitions: an initial transition into an $s_1+s_2$ state followed by a second transition into the BTRS state, $s_1+is_2$\cite{Maiti2013}. To date, no experimental evidence for such a double transition has been reported, making this scenario unlikely.

A natural consequence of the chiral $d+id$ superconducting state suggested by our results is BTRS. Although measurements that directly probe BTRS have not yet been reported for hole-doped RbV$_3$Sb$_{5-x}$Sn$_x$, $\mu$SR measurements performed under hydrostatic pressure reveal a remarkably similar phenomenology. Increasing pressure simultaneously suppresses the charge-density wave, enhances $T_c$, and promotes a fully gapped superconducting state. In this regime, the $\mu$SR relaxation rate shows an enhancement below $T_c$, a feature that has been commonly associated with BTRS superconductivity \cite{guguchia2023tunable}. Furthermore, no impurity-induced bound states were observed in samples at lower doping, suggesting either that unconventional superconductivity develops only near optimal hole doping or that its identification by STM requires suitable intrinsic defects.

\section{Methods}\label{sec:Methods}
\subsection{Theory}
Here we introduce a theoretical model used to simulate the $dI/dV$-curve at the impurity site and the spatial profile.

\subsubsection{Model}
\begin{figure}[!ht]
\centering
\includegraphics[width=0.5\linewidth]{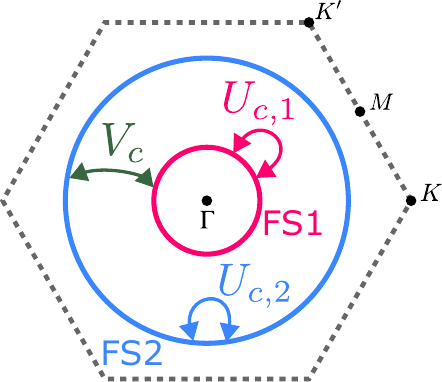}
\caption{Schematic description of the two Fermi surfaces. The inner pink Fermi surface is dominated by planar Sb states, whereas the outer blue Fermi surface contains states from V orbitals and apical Sb. The couplings $U_{c,1}$ and $U_{c2}$ represent intra-Fermi surface scattering, whereas $V_c$ is an inter-Fermi surface scattering. These couplings are defined mathematically in Eq.~\eqref{eq:VcBW}.}
\label{fig:twoFSsSchmatic}
\end{figure}

To model Imp-BS in ${\rm RbV_3Sb_5}$, we consider a multi-band system featuring two Fermi surfaces, as schematically illustrated in Fig.~\ref{fig:twoFSsSchmatic}.

Assuming spin-singlet superconductivity and neglecting spin-orbit coupling, the system is described by the following Bogoliubov-de Gennes (BdG) Hamiltonian:
\begin{align}
H_{\rm BdG}(\mathbf{k}) &= \Bigg[\Big(E_1(\mathbf{k})\tau_3 - {\rm Re}\Delta_1(\mathbf{k})\tau_1 + {\rm Im}\Delta_1(\mathbf{k})\tau_2\Big)\otimes \frac{\mathbb{1}+\eta_3}{2} \nonumber \\
&\quad + \Big(E_2(\mathbf{k})\tau_3 - {\rm Re}\Delta_2(\mathbf{k})\tau_1 + {\rm Im}\Delta_2(\mathbf{k})\tau_2\Big)\otimes \frac{\mathbb{1}-\eta_3}{2} \Bigg]\otimes \sigma_0,
\label{eq:BdGHam}
\end{align}
where the subscripts $1$ and $2$ denote the two Fermi surfaces (FS1 and FS2), respectively, shown in Fig.~\ref{fig:twoFSsSchmatic}. The matrices $\sigma_i$ and $\tau_i$ represent Pauli matrices acting in spin and particle-hole spaces, while $\eta_i$ ($i=1,2,3$) act in the Fermi-surface (orbital) space. For a general spin-singlet gap, we consider both the real and imaginary components of the superconducting order parameter on each Fermi surface.

The total Hamiltonian is expressed as $\hat{H} = \sum_{\mathbf{k}} \Psi^\dagger(\mathbf{k}) H_{\rm BdG}(\mathbf{k}) \Psi(\mathbf{k})$, where $\Psi(\mathbf{k})$ is the Balian-Werthammer (BW) spinor defined by
\begin{align}
\Psi(\mathbf{k}) = \begin{pmatrix}
\psi_{1,\uparrow}(\mathbf{k}) &
\psi_{1,\downarrow}(\mathbf{k}) &
\psi_{2,\uparrow}(\mathbf{k}) &
\psi_{2,\downarrow}(\mathbf{k}) &
-\psi^\dagger_{1,\downarrow}(-\mathbf{k}) &
\psi^\dagger_{1,\uparrow}(-\mathbf{k}) &
-\psi^\dagger_{2,\downarrow}(-\mathbf{k}) &
\psi^\dagger_{2,\uparrow}(-\mathbf{k})
\end{pmatrix}^T.
\label{eq:TwoFSNBspinor}
\end{align}
 The charge (i.e., nonmagnetic) and magnetic impurity potentials are modeled as
\begin{align}
\hat{V}_c &= \sum_{\alpha=1,2}\sum_{\sigma=\uparrow,\downarrow} U_{c,\alpha} \psi^\dagger_{\alpha,\sigma}(\mathbf{r}=0)\psi_{\alpha,\sigma}(\mathbf{r}=0) \nonumber \\
&\quad + V_c \sum_{\sigma=\uparrow,\downarrow} \Big(
\psi^\dagger_{1,\sigma}(\mathbf{r}=0)\psi_{2,\sigma}(\mathbf{r}=0) + 
\psi^\dagger_{2,\sigma}(\mathbf{r}=0)\psi_{1,\sigma}(\mathbf{r}=0)
\Big),\\
\hat{V}_m &= \sum_{\alpha=1,2}\sum_{\sigma=\uparrow,\downarrow} \sigma U_{m,\alpha} \psi^\dagger_{\alpha,\sigma}(\mathbf{r}=0)\psi_{\alpha,\sigma}(\mathbf{r}=0) \nonumber \\
&\quad + V_m \sum_{\sigma=\uparrow,\downarrow} \sigma \Big(
\psi^\dagger_{1,\sigma}(\mathbf{r}=0)\psi_{2,\sigma}(\mathbf{r}=0) + 
\psi^\dagger_{2,\sigma}(\mathbf{r}=0)\psi_{1,\sigma}(\mathbf{r}=0)
\Big),
\end{align}
where $U_{c(m),\alpha}$ ($\alpha=1,2$) represents intra-Fermi-surface scattering, and $V_{c(m)}$ accounts for inter-Fermi-surface scattering processes. In the BW basis, the impurity potential in momentum space is given by $\hat{V}_{c(m)} = \sum_{\mathbf{k},\mathbf{k}'} \Psi^\dagger(\mathbf{k}) V_{c(m)}^{(\text{BW})} \Psi(\mathbf{k}')$, where
\begin{align}
V_c^{(\text{BW})} &= \tau_3\otimes \Bigg[
\frac{U_{c,1}+U_{c,2}}{2}\eta_0 + 
\frac{U_{c,1}-U_{c,2}}{2}\eta_3 + 
V_c\eta_1
\Bigg]  \otimes \sigma_0,
\label{eq:VcBW}\\
V_m^{(\text{BW})} &= \tau_0\otimes\Bigg[
\frac{U_{m,1}+U_{m,2}}{2}\eta_0 + 
\frac{U_{m,1}-U_{m,2}}{2}\eta_3 + 
V_m\eta_1
\Bigg]  \otimes \sigma_3.
\label{eq:VmBW}
\end{align}

\subsubsection{Local Density of States (LDOS) of impurity bound states}
The $dI/dV$ curve measured in STM is given as follows:
\begin{align}
     \frac{dI}{dV}(V,\mathbf{r},T)&\propto \int dE \mathcal{N}(E,\mathbf{r})K(E-eV,T)
\end{align}
where $K(x,T)=-\frac{\partial f_{\rm FD}(x,T)}{\partial x}$ and $\mathcal{N}(E,\mathbf{r})$ is the local density of states (LDOS) at energy $E$ and position $\mathbf{r}$, given by:
\begin{align}
    \mathcal{N}(E,\mathbf{r}) &= -2\sum_{\alpha=1,2}\sum_{\sigma=\uparrow,\downarrow} {\rm Im} \left[ -\langle \psi_{\alpha,\sigma}(i \omega_n, \mathbf{r}) \psi_{\alpha,\sigma}^\dagger(i \omega_n, \mathbf{r}) \rangle \Big|_{i \omega_n \rightarrow E+i\eta} \right] \nonumber \\
    &= -2 {\rm ImTr} \left[ \Big( G(i \omega_n \rightarrow E+i\eta, \mathbf{r}, \mathbf{r}) \Big)_{pp} \right], \label{eq:LDOS}
\end{align}
where $G(i \omega_n, \mathbf{r}, \mathbf{r})$ is the full BdG Green function including impurity scattering, and the subscript $pp$ denotes the particle-particle (electron-like) sector extracted from the full Green function.

Before deriving the explicit form of the full Green function, we address several key assumptions in Eq.~\eqref{eq:LDOS}. In principle, the LDOS should be represented in the spin, orbital, and sublattice basis rather than the band (Fermi surface) basis. However, in the kagome system under consideration, several factors justify our approach. First, the spin-orbit coupling is weak\cite{bhandari_first-principles_2024}. Second, the two Fermi surfaces exhibit strong sub-lattice selectivity~\cite{ritz2023superconductivity, schultz_superconductivity_2026}: the small Fermi surface near the $\Gamma$ point consists primarily of Sb electrons, while the large Fermi surface near the Brillouin zone boundary is dominated by V electrons. Consequently, we expect minimal differences between the band basis and the orbital/sublattice basis.

The full Green function $G(i \omega_n, \mathbf{r}, \mathbf{r})$ in the presence of an impurity potential is given by:
\begin{align}
    G(i \omega_n, \mathbf{r}, \mathbf{r}) &= G_0(i \omega_n, \mathbf{r}=0) + G_0(i \omega_n, \mathbf{r}) T(i \omega_n) G_0(i \omega_n, -\mathbf{r}), \\
    T(i \omega_n) &= \left( \mathbb{1} - V_{c(m)}^{(\text{BW})} G_0(i \omega_n, \mathbf{r}=0) \right)^{-1} V_{c(m)}^{(\text{BW})}, \label{eq:TMatrix}
\end{align}
where $G_0(i \omega_n, \mathbf{r})$ is the bare Green function obtained from Eq.~\eqref{eq:BdGHam}, $T(i \omega_n)$ is the T-matrix, and $V_{c}^{(\text{BW})}$ is the impurity potential matrix from Eq.~\eqref{eq:VcBW}. The explicit form of the bare BdG Green function in momentum space is:
\begin{align}
    G_0(i \omega_n, \mathbf{k}) &= \Bigg[ G_{1,0}(i\omega_n,\mathbf{k}) \frac{\mathbb{1}+\eta_3}{2} + G_{2,0}(i\omega_n,\mathbf{k})\frac{\mathbb{1}-\eta_3}{2} \Bigg] \otimes \sigma_0,\label{eq:G0inMomentum}\\
    G_{\alpha,0}(i\omega_n,\mathbf{k})&=-\frac{i \omega_n\tau_0 + E_\alpha(\mathbf{k})\tau_3 + {\rm Re}\Delta_\alpha(\mathbf{k})\tau_1 - {\rm Im}\Delta_\alpha(\mathbf{k})\tau_2}{\omega_n^2 + (E_\alpha(\mathbf{k}))^2 + |\Delta_\alpha(\mathbf{k})|^2}
\end{align}
where $\Delta_\alpha(\mathbf{k}) = {\rm Re}\Delta_\alpha(\mathbf{k}) + i{\rm Im}\Delta_\alpha(\mathbf{k})$ and $\alpha(=1,2)$ is a Fermi surface index.

The spectrum of impurity-induced bound states is determined by the condition ${\rm Re\Big[ det}\left(T^{-1}(i\omega_n\rightarrow \omega+i\eta)\right)\Big]\propto {\rm \Big[det}\left( \mathbb{1} - V_{c(m)}^{(\text{BW})} G_0(\omega+i\eta, \mathbf{r}=0) \right)\Big] = 0$ within the superconducting gap range $|\omega| < |\Delta|$. As seen from Eq.~\eqref{eq:TMatrix}, these bound state solutions depend solely on the local Green function $G_0(i \omega_n, \mathbf{r}=0)$. Thus, the energy positions of the bound states remain constant regardless of the distance from the impurity site. While particle-hole symmetry ensures pairs of solutions at $\omega=\pm E_{\rm Imp-BS}$, this does not imply that two symmetric peaks will always appear in the LDOS at a general position $\mathbf{r}$. This asymmetry arises because the LDOS selectively probes the particle-particle sector (Eq.~\eqref{eq:LDOS}), providing a crucial theoretical tool to distinguish gap symmetries. 

To obtain the LDOS, it is necessary to calculate the Green function in position space from Eq.~\eqref{eq:G0inMomentum}. For each Fermi surface, the Green function in position space is given by
\begin{align}
    G_{\alpha,0}(i\omega_n,\mathbf{r})&=\int \frac{d^2\mathbf{k}}{(2\pi)^2}G_{\alpha,0}(i\omega_n,\mathbf{k})e^{i\mathbf{k}\cdot\mathbf{r}}\nonumber\\
    &=-\int\frac{d^2\mathbf{k}}{(2\pi)^2}\frac{i \omega_n\tau_0 + E_\alpha(\mathbf{k})\tau_3 + {\rm Re}\Delta_{\alpha}(\mathbf{k})\tau_1 - {\rm Im}\Delta_{\alpha}(\mathbf{k})\tau_2}{\omega_n^2 + (E_\alpha(\mathbf{k}))^2 + |\Delta_{\alpha}(\mathbf{k})|^2}\otimes \sigma_0 e^{i\mathbf{k}\cdot\mathbf{r}}\nonumber\\
    &\approx -\int_{-D}^{D}d\epsilon \rho_\alpha(\epsilon)\int \frac{d\theta_{k}}{2\pi}e^{i\left(k_{F,\alpha}+\frac{\epsilon}{\hbar v_{F,\alpha}}\right)r\cos(\theta_k-\phi_r)}\nonumber\\
    &\times \frac{i \omega_n\tau_0 + \epsilon\tau_3 + {\rm Re}\Delta_{\alpha}(\theta_k)\tau_1 - {\rm Im}\Delta_{\alpha}(\theta_k)\tau_2}{\omega_n^2 + \epsilon^2 + |\Delta_{\alpha}(\theta_k)|^2}\otimes \sigma_0 
\end{align}
where subscript $\alpha(=1,2)$ denotes the Fermi surface, and $\theta_k$ and $\phi_r$ are the angles of the wave vector $\mathbf{k}$ and position vector $\mathbf{r}$, respectively. $\rho_\alpha(\epsilon)$, $k_{F,\alpha}$ and $v_{F,\alpha}$ denote the density of states, Fermi momentum and Fermi velocity of Fermi surface $\alpha$, respectively.

In the discussion section, we discussed possible sources of particle-hole asymmetric YSR states on magnetic impurities. There, we discussed the simultaneous charge and magnetic impurity potential, and also the Fano line shape from the Kondo effect. A further possibility is that an asymmetry already exists in the normal state electronic density of states, and is inherited by the YSR peaks. Hence, to allow the possibility of an asymmetric density of states, we use the following form for $\rho_\alpha(\epsilon)$:
\begin{align}
    \rho_\alpha(\epsilon)=\rho_{\alpha,0}\left[1+\lambda_\alpha \tanh\left(\frac{\epsilon}{\Lambda_\alpha}\right)\right],
\end{align}
where $\rho_{\alpha,0}$ is the density of states of Fermi surface $\alpha$ at the Fermi energy and $\lambda_\alpha < 1$ denotes the magnitude of the asymmetry described by the $\tanh$ function. $\Lambda_\alpha$ characterizes the width of the asymmetry. 

With the assumption that size of the gap does not change along the Fermi line, we can assume the following form of the gap  
\begin{align}
    \Delta_{\alpha}(\mathbf{k})={\rm Re}\Delta_{\alpha}(\mathbf{k})+i{\rm Im}\Delta_{\alpha}(\mathbf{k})\approx 
    -\Delta_{\alpha,0}e^{in\theta_{k}},\;  \begin{cases}
        n=0  & \text{$s$-wave gap}\\
        n=2 & \text{$d+id$-wave gap}
    \end{cases}
\end{align}
where $\Delta_{\alpha,0}$ is the $\theta_k$-independent part of the gap. In the case of the $s$-wave gap structure, the relative sign difference between $\Delta_{1,0}$ and $\Delta_{2,0}$ determines whether it has $s^{\pm}$ or $s^{++}$ gap structures.

With the above assumptions, we can obtain the following general form of $G_{\alpha,0}(\omega+i\eta,\mathbf{r})$ through the analytical continuation $i\omega_n\rightarrow \omega+i\eta$:
\begin{align}
    G_{\alpha,0}(\omega+i\eta,\mathbf{r})&=\begin{pmatrix}
        g_{\alpha}(\omega+i\eta,\mathbf{r})+h_\alpha(\omega+i\eta,\mathbf{r}) & f_\alpha(\omega+i\eta,\mathbf{r}) \\ 
        f_\alpha^*(\omega+i\eta,\mathbf{r}) & g_{\alpha}(\omega+i\eta,\mathbf{r})-h_\alpha(\omega+i\eta,\mathbf{r})
    \end{pmatrix}\\
g_\alpha(\omega,\mathbf{r})&=-\int_{-D}^D d\epsilon \rho_\alpha (\epsilon) \frac{\omega+i\eta}{-(\omega+i\eta)^2+\epsilon^2+\Delta_{\alpha,0}^2}J_0\left(k_{F,\alpha}r+\frac{\epsilon r}{\pi\xi_{\alpha}|\Delta_{\alpha,0}|}\right),\\
h_\alpha(\omega,\mathbf{r})&=-\int_{-D}^D d\epsilon \rho_\alpha(\epsilon) \frac{\epsilon}{-(\omega+i\eta)^2+\epsilon^2+\Delta_{\alpha,0}^2}J_0\left(k_{F,\alpha}r+\frac{\epsilon r}{\pi\xi_{\alpha}|\Delta_{\alpha,0}|}\right),\\
f_\alpha(\omega,\mathbf{r})&=\Delta_{\alpha,0} e^{2i\phi_r}\int_{-D}^D d\epsilon \rho_\alpha(\epsilon)\frac{1}{-(\omega+i\eta)^2+\epsilon^2+\Delta_{\alpha,0}^2}\nonumber\\
&\times \begin{cases}J_0\left(k_{F,\alpha}r+\frac{\epsilon r}{\pi\xi_{\alpha}|\Delta_{\alpha,0}|}\right) & \text{$s$-wave gap}\\
J_2\left(k_{F,\alpha}r+\frac{\epsilon r}{\pi\xi_{\alpha}|\Delta_{\alpha,0}|}\right) & \text{$d+id$-wave gap}\end{cases}
\end{align}
where $\xi_\alpha=\frac{\hbar v_{F,\alpha}}{\pi|\Delta_{\alpha,0}|}$. Using Eq.~\eqref{eq:LDOS} with a charge impurity potential Eq.~\eqref{eq:VcBW} and the numerically obtained $g_\alpha$, $h_\alpha$ and $f_\alpha$ functions, we obtain the results presented in Fig.~\ref{fig:defect_type1} and \ref{fig:defect_type2}.

We can quantify the peak-height asymmetry of the two Imp-BS peaks at $\pm E_{\rm Imp-BS}$ in terms of the $g_{\alpha}(\omega,\mathbf{r})$ and $f_\alpha(\omega,\mathbf{r})$ as follows:
\begin{align}
    Z_+= |g_\alpha(\omega=E_{\rm Imp-BS},\mathbf{r}=0)|^2,\quad Z_-=|f_\alpha(\omega=-E_{\rm Imp-BS},\mathbf{r}=0)|^2\label{eq:DefZ}
\end{align}
where the peak heights at $E_{\rm Imp-BS}$ and $-E_{\rm Imp-BS}$ proportional to $Z_+$ and $Z_-$ respectively. From the above definition of $Z_-$, we can also get the explicit formula given in Eq.~\eqref{eq:Zminums}.

\subsubsection{Numerical parameters}
For the theoretical calculations of both impurity cases, we adopt the following parameters for the density of states:
\begin{align}
    \frac{\rho_{0,1}}{\rho_{0,2}}=2,\quad \lambda_{1}=0.3,\quad \lambda_2=0.2,\quad \Lambda_1=\Lambda_2=2\Delta_1.
\end{align}

For the spatial profile in Fig.~\ref{fig:defect_type1}, the coherence lengths are chosen as:
\begin{align}
    \frac{\xi_1}{a}=1.337,\quad \frac{\xi_2}{a}=10.1
\end{align}
where $a = 0.55\,\text{nm}$ is the lattice constant. Here, the values of $\xi_1$ and $\xi_2$ are determined using the relation $\xi_\alpha\propto \frac{1}{|\Delta_{\alpha,0}|\rho_{\alpha,0}}$.

The following fixed parameters are obtained from the experimental data:
\begin{align}
    k_{F,1}=1.79\,\text{nm}^{-1},\quad k_{F,2}=7.24\,\text{nm}^{-1},\quad \Delta_{1,0}=0.68\,\text{meV},\quad \Delta_{2,0}=0.18\,\text{meV}.
\end{align}
Finally, the impurity potentials (dimensionless since they are multiplied by the DOS at the Fermi energy) are set to:
\begin{align}
    (U_{c,1},U_{c,2},V_c)=\begin{cases}
        (-0.2,-0.25,0.25) & \text{Defect 1}\\
        (-0.23,0,0.05) & \text{Defect 2}
    \end{cases}.
\end{align}

\subsection{STM}
STM measurements were performed using a Unisoku STM head at an instrument temperature of 0.27~K, using chemically etched and annealed tungsten tips. Spectra were acquired using a standard lock-in technique at a frequency of 907.7 Hz. 

\section{Acknowledgements}
We are grateful to Andreas Kreisel, and Brian M{\o}ller Andersen for stimulating discussions. 

\section{Funding}
The experimental work at UIUC was supported by a grant from the US Department of Energy, Office of Science, Basic Energy Sciences, under award number DE-SC0022101. A.A acknowledge support from the US National Science Foundation (NSF) GrantNumber 2201516 under the Accelnet program of Office of International Science and Engineering (OISE). The work at KIT was supported by the German Research Foundation (DFG) through CRC TRR 288 ``Elasto-Q-Mat,'' project A07 (I.J., D.J.S. and J.S.), and the Simons Foundation Collaboration on New Frontiers in Superconductivity (Grant SFI-MPS-NFS-00006741-03) (J.S.). R.M.F.  acknowledges a Mercator Fellowship from the German Research Foundation (DFG) through CRC TRR 288, 422213477 ``Elasto-Q-Mat.'' Funding for sample growth was provided via the UC Santa Barbara NSF Quantum Foundry funded via the Q-AMASE-i program under award DMR-1906325. A.N.C.S. acknowledges support from the Eddlemam Center for Quantum Innovation at UC Santa Barbara.

\section {Author contribution}
A.A., Y.X. and V.M. conceived the project. The single crystals were provided by A.N.C. and S.D.W. A.A. and Y.X. obtained the STM data. A.A. and V.M. performed the analysis and I.J., D.J.S., G.P., R.M.F. and J.S. provided theoretical calculations on the interpretation of the data. A.A., V.M., D.J.S., I.J., J.S. and R.M.F. wrote the paper with input from all authors.

\section{Competing Interest Statement}

The authors declare no competing interests.
\clearpage

\bibliography{Mainbib}

\clearpage

\begin{titlepage}
	\centering
	\vspace*{1cm}
	
	{\Huge Supplementary Information for Direct Evidence of Unconventional Superconductivity in Doped Kagome system RV$_3$Sb$_5$}\\[0.5cm]
	
	\vfill
\end{titlepage}

\clearpage
	
	\setcounter{section}{0}
	\setcounter{figure}{0}
	\renewcommand*{\figurename}{\textbf{Supplementary Figure}}
	\pagebreak
	\section{Cleaning the surface from Rb atoms}
	
	As in any AV$_3$Sb$_5$ STM study, when the sample is cleaved, the cleave plane is always between the alkali metal layer and the Sb honeycomb layer. Residual alkali adatoms remain on the Sb layer. Thus, consistent with previous STM studies\cite{jiang2021unconventional,chen2021roton,xing2024optical,di2026kagome,zhao2021cascade,li2023unidirectional,li2022rotation,nie2022charge}, we clean the sample by mechanically moving the adatoms to the edges of the scanning area using the tungsten tip. This process exposes a clean Sb surface for tunneling experiments, as seen in Supplementary Figure S1.
	
	\begin{figure}[!ht]
		\includegraphics[trim= 0cm 7cm 0cm 4cm,clip=true,width=1\textwidth]{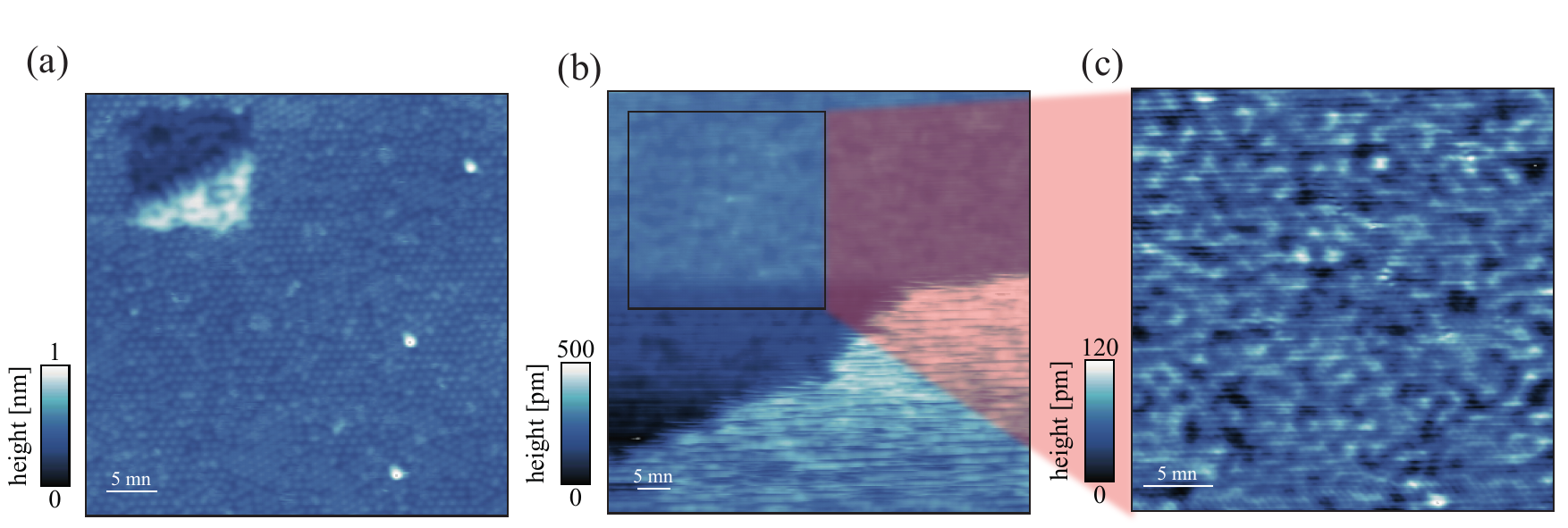}
		\caption{\textbf{Cleaved surface before and after cleaning} \textbf{a} exposed Rb surface, after cleaving at 77K. The atoms are well separated, with some observed reconstruction of the Rb layer. \textbf{b} Exposed Sb layer after cleaning the surface from Rb atoms. The atoms were pushed downwards to create the aggregated layer at the bottom of the scan. The Black square marks the location of the scan in c. \textbf{c} Exposed Sb honeycomb layer after cleaning the Rb atoms.  (\textbf{a} $V_{\text{Bias}}=-90$~mV, $I_{\text{Set}} =10$~pA. \textbf{b} $V_{\text{Bias}}=-10$~mV, $I_{\text{Set}}=100$~pA. \textbf{c} $V_{\text{Bias}}=-10$~mV, $I_{\text{Set}}=150$~pA. }
		
		\label{FigS1}
	\end{figure}
	\pagebreak
	\section{Gap dependence on temperature and magnetic field}
	
	\begin{figure}[!ht]
		\includegraphics[trim= 0cm 3cm 0cm 3cm,clip=true,width=1\textwidth]{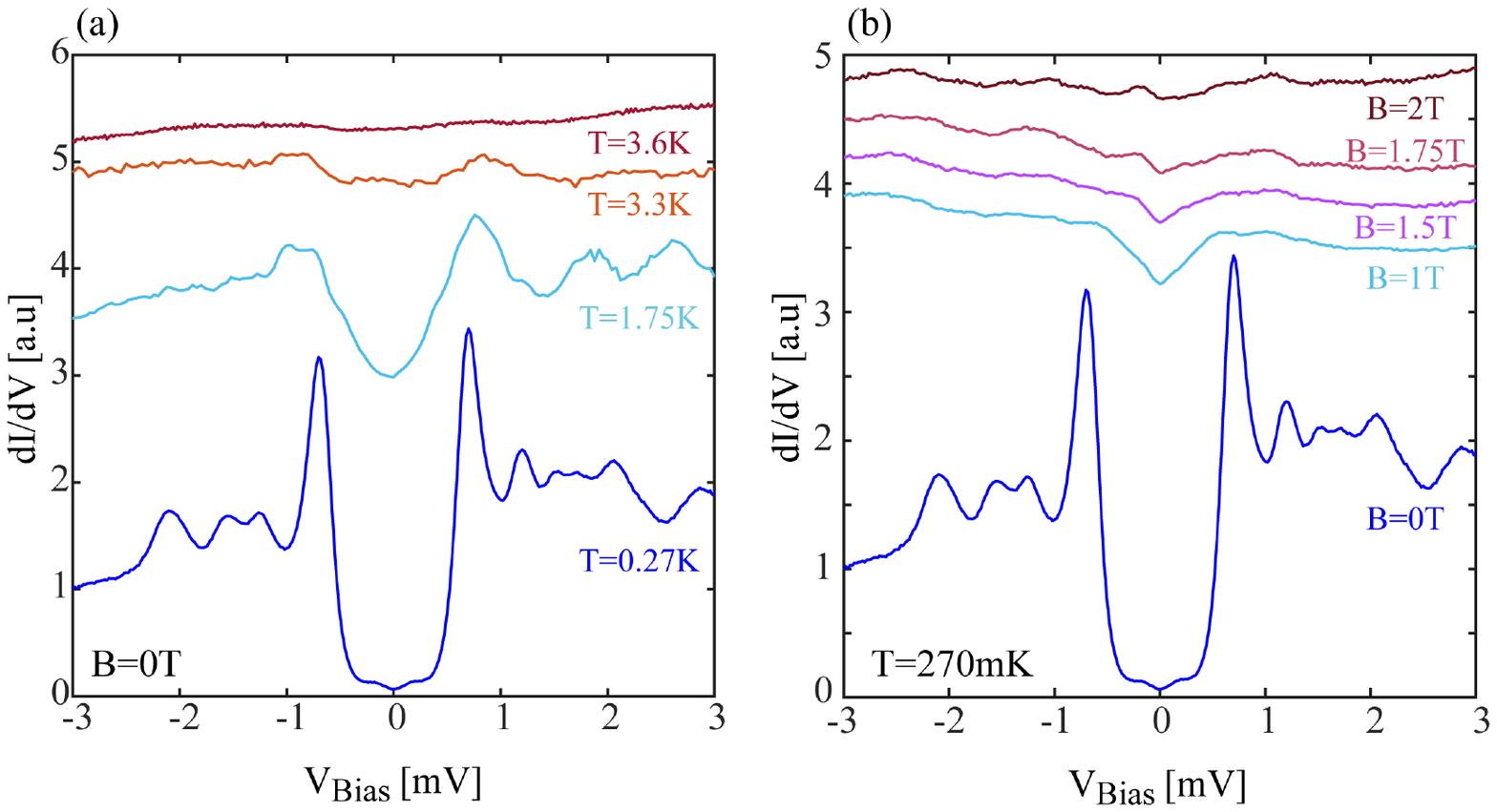}
		\caption{\textbf{Temperature and field dependence of the superconducting gap.} \textbf{a,b} Temperature and magnetic field dependence of the superconducting field, respectively. The small coherence peaks inside the big gap are absent already at 1.75~K. The gap is completely filled at 3.6K, which match $T_{C}$ from transport measurements. Similarly, the gap is filled at 2 T, which gives us $H_{c2}=2$T. Both data sets were acquired at the same location shown in Fig.~1. $V_{\text{Set}}=-3$ mV, $I_{\text{set}}=150$ pA, $V_{\text{Modulation}}=30\mu$V}
		
		\label{FigS2}
	\end{figure}
	
	\pagebreak
	\section{Doping in the Kagome layer and associated band structure change}
	
	To characterize the locations of the Sn dopants, we start by imaging isolated Sn atoms in lightly doped samples (Supplementary Fig.~3). The observed sixfold symmetry reflects the coordination of the substituted Sb atom by six nearest-neighbor V atoms, in contrast to the threefold coordination expected for substitution within the Sb honeycomb layer. The symmetry demonstrates that Sn substitutes the planar Sb atoms within the kagome layer rather than the Sb atoms in the honeycomb termination, marked as Sb~1 sites in Fig.~1 b. Furthermore, The reduced topographic contrast at the dopant center indicates a locally suppressed density of states, consistent with hole doping. 
	
	Independent evidence for substitution within the kagome layer is obtained from the evolution of the Friedel oscillations. In AV$_3$Sb$_5$, the dominant Friedel oscillations originate from scattering within the planar Sb-derived Fermi-surface pocket centered at $\Gamma$\cite{shumiya2021intrinsic,liang2021three,di2026kagome,zhao2021cascade,li2022rotation,wang2021charge}. The corresponding scattering wave vector, extracted from the FFT of the topography, decreases systematically with increasing Sn concentration (Fig.~1 c,d and Supplementary Fig.~4), demonstrating that the $\Gamma$ pocket shrinks upon hole doping. This behavior agrees quantitatively with both ARPES and DFT. ARPES measurements report $(k_F)_{\rm ARPES} = 2.11 \pm 0.125~\mathrm{nm}^{-1}$ for the parent compound\cite{kato2023surface}, corresponding to $(q_{\rm Friedel})_{\rm ARPES} = 4.22 \pm 0.25 ~\mathrm{nm}^{-1}$, while DFT predicts a reduction of $(q_{\rm Friedel})_{\rm DFT}$ from $3.95~\mathrm{nm}^{-1}$ ($x=0$) to $3.64~\mathrm{nm}^{-1}$ ($x=0.06$)\cite{oey2022tuning}. Our measurements yield $q_{\rm Friedel}=4.02 \pm 0.19~\mathrm{nm}^{-1}$ and $3.58\pm0.19~\mathrm{nm}^{-1}$ for $x=0$ and $x=0.06$, respectively, in good agreement with these predictions. Together with the real-space imaging, these results demonstrate that Sn predominantly substitutes the planar Sb sites within the kagome layer. Our conclusions are further supported by nuclear quadrupole resonance measurements on CsV$_3$Sb$_{5-x}$Sn$_x$, which likewise identify preferential substitution of the planar Sb sites\cite{k2026observation,oey2022fermi}. 
	
	Density functional theory (DFT) calculations indicate that full substitution of the Sb atom at the center of the Kagome Star-of-David cluster with Sn shifts the Sb-derived band near the $\Gamma$ and A points of the Brillouin zone upward by approximately 1 eV\cite{oey2022tuning}. The hole doping caused by the Sn atoms results in a shift of the V-derived bands to higher energies by about 200~meV, compared to the parent compound. By interpolating between the DFT calculations for RbV$_3$Sb$_5$ and RbV$_3$Sb$_4$Sn\cite{oey2022tuning}, overall we find that for partial doping level of $x=0.3$, the system is expected to exhibit an intermediate behavior, combining a rigid shift of the Sb bands with a chemical potential shift of the V bands. The primary effect of hole doping is a redistribution of the bandwidth among the different orbital manifolds: the planar Sb-$p$ pocket at $\Gamma$ contracts by as much as 0.3 eV, and the V-$d$ derived bands shift such that the Van-Hove singularity at the M point approaches the Fermi level, remaining only a few tens of meV below $E_F$.
	
	\begin{figure}[!ht]
		\includegraphics[trim= 0cm 2cm 0cm 3cm,clip=true,width=1\textwidth]{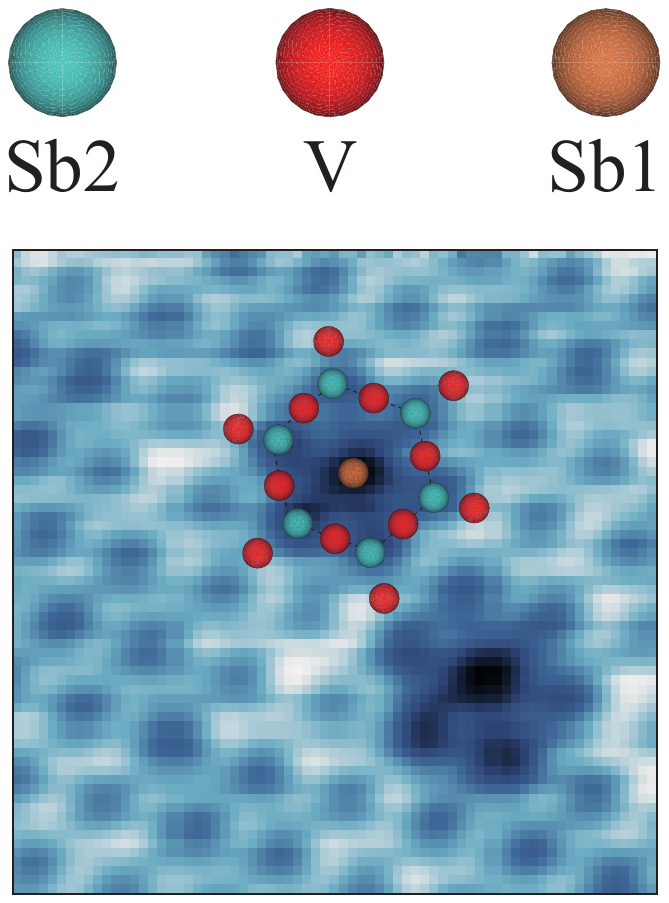}
		\caption{\textbf{Topography of isolated dopant.} Topography scan of two isolated Sn doping sites shown as a six fold symmetric "flower". Data was taken on an Sb honeycomb layer on a $\rm{RbV_3Sb_{4.99}Sn_{0.01}}$. In both cases the Sb honeycomb layer is visible and the doping site is in the center of the honeycomb unit cell, suggesting it to be in the center of the Kagome unit cell, where the Sb atom resides in the V-Sb layer. (V$_S$=-50 mV, I$_t$=50 pA.)}
		
		\label{FigS3}
	\end{figure}
	
	\begin{figure}[!ht]
		\includegraphics[trim= 0cm 0cm 0cm 0cm,clip=true,width=1\textwidth]{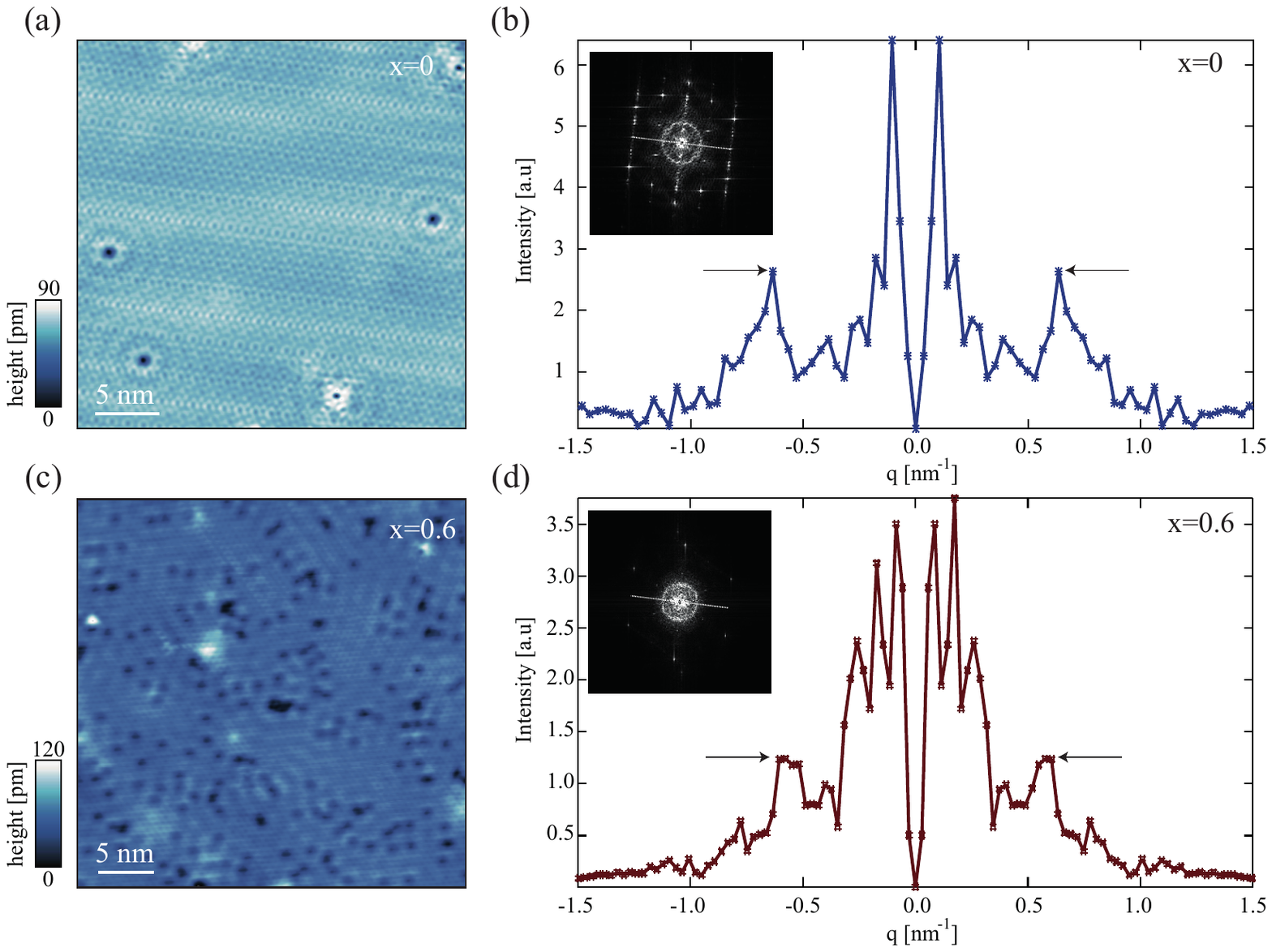}
		\caption{\textbf{The evolution of Friedel oscillations with doping.} \textbf{a} Topography of the Sb honeycomb layer in a parent compound sample. Holes in the surface are visible, which creates the Friedel oscillations, seen as ripples around each hole. \textbf{b} Intensity along a line-cut taken from the FFT of the topography shown in \textbf{a}. Black arrows mark the location of peaks corresponding to the Friedel oscillation q vector. Inset: FFT of the topography shown in \textbf{a}. White dashed line marks the line-cut. \textbf{c} Topography of the Sb honeycomb layer in doped sample ($x=0.06$). The Friedel oscillations are masked by the change in the density of states created by the Sn doping. \textbf{d} Same as \textbf{b} for the topography shown in \textbf{c}. Black arrows mark the location of peaks corresponding to the Friedel oscillation q vector. Inset: FFT of the topography shown in \textbf{c}. White dashed line marks the line-cut.  (\textbf{a} $V_S=-10$ mV, $I_t=250$ pA. \textbf{b} $V_S=-30$ mV, $I_t=250$ pA.)}
		
		\label{FigS4}
	\end{figure}
	
	\clearpage
	\section{Superconducting conductance maps in slightly doped and parent compounds}
	
	Here we show both topography and $E=0$ meV energy slice taken from $dI/dV$ maps acquired on other doping levels. The $dI/dV$ maps lack any signature of bound states, as discussed in the main text. This is shown in Supplementary Figure S7 for both the parent compound ($x=0$) and slightly doped sample ($x=0.06$). We note that the lack of signature is also for energies away from the Fermi level, and with-in the superconducting gap. It is visible that both the Sn doping and missing atoms in the Sb honeycomb layer are not affecting the superconducting spectra, thus are not responsible for the appearance of the bound states. 
	
	\begin{figure}[!ht]
		\includegraphics[trim= 0cm 0cm 0cm 0cm,clip=true,width=1\textwidth]{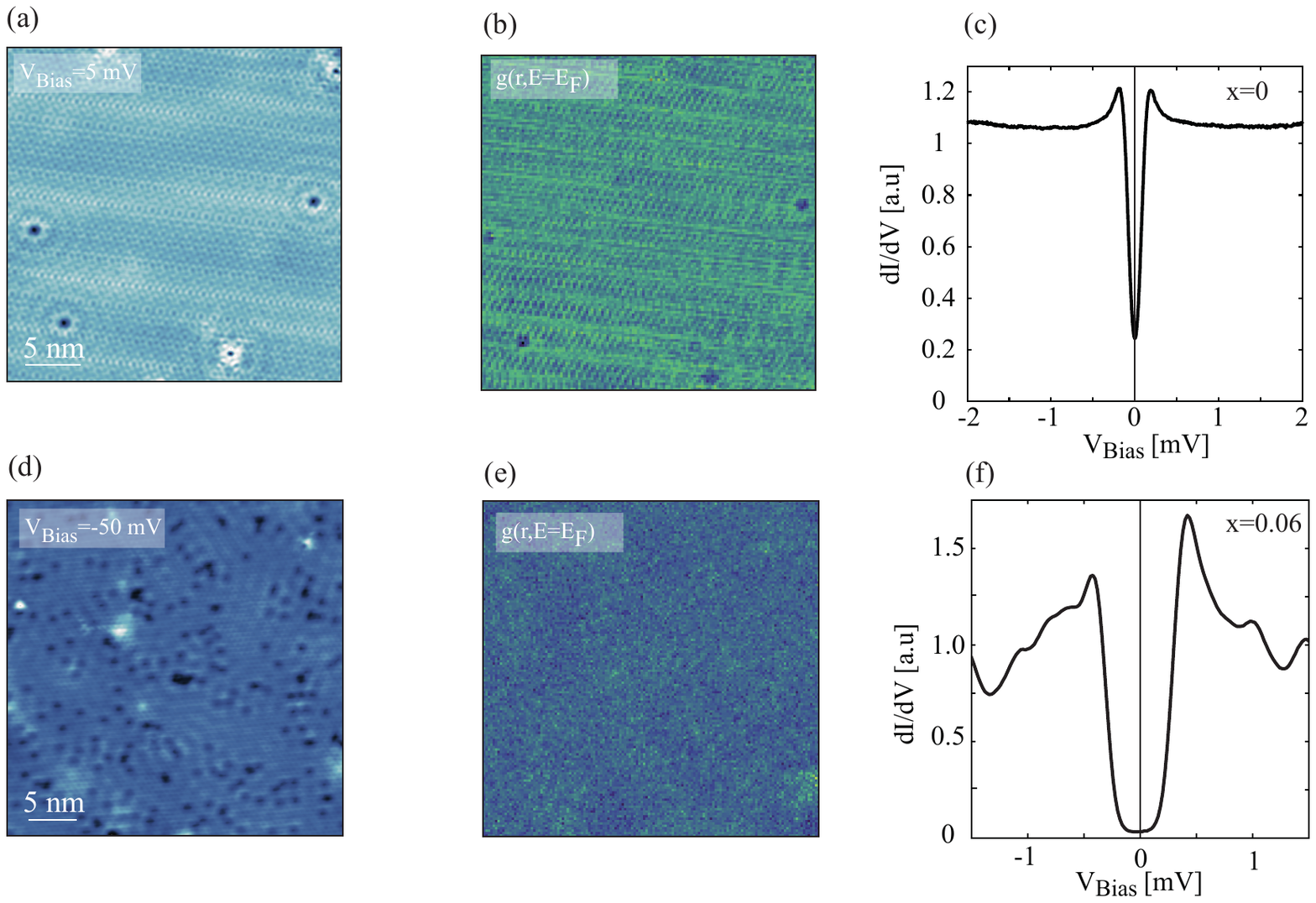}
		\caption{\textbf{Conductance maps at the Fermi level for different doping Levels.} \textbf{a,b} Conductance maps out side (5 meV) and at the Fermi level ($E_f$) of the pristine compound of \RVS\ ($x=0$). Local holes in the Sb honeycomb lattice are visible, producing Friedel oscillations which we use to measure $k_F$ of the Sb derived band.  $V_{\text{Set}}=-5$~mV, $I_{\text{Set}}=~200$ pA. \textbf{c} Superconducting spectrum of the same sample, acquired as an average of multiple spectra, all of which measured in the are shown in a and b. $V_{\text{Set}}=2$~mV, $I_{\text{Set}}=~150$ pA. \textbf{d,e,f} Same for \RVS\ ($x=0.06$). for f: $V_{\text{Set}}=-4$~mV, $I_{\text{Set}}=~300$ pA. a-f temperature is $T=0.27$ K.}
		\label{FigS7}
	\end{figure}

	\clearpage
	
	\section{System calibration}
	We use pure Pb to calibrate the effective temperature of the system prior to measuring superconductors. This is done to ensure minimal noise interference is incorporated in the signal of small superconducting gaps, as seen in the Supplementary Figure~\ref{FigS7}. 
	
	\begin{figure}[!ht]
		\includegraphics[trim= 0cm 3cm 0cm 2cm,clip=true,width=1\textwidth]{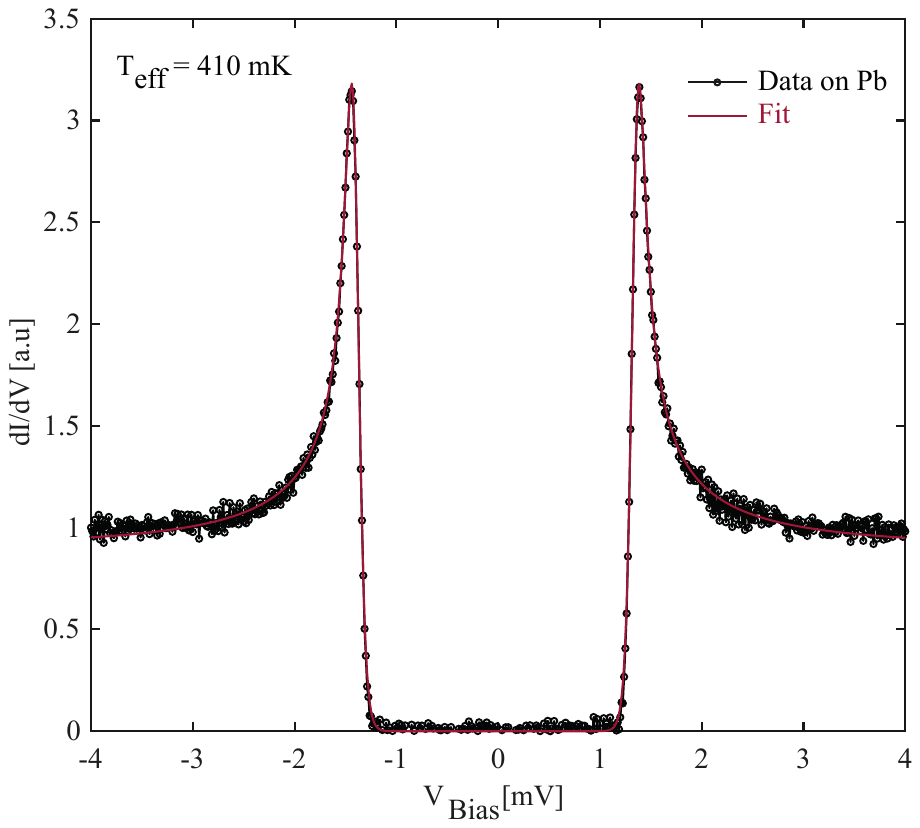}
		\caption{\textbf{Calibration of the effective temperature with Pb.} Superconducting gap measured on the surface of pure Pb. The fit was done using a BCS function which gives the low effective temperature of the system (with nominal temperature of 270~mK). Measurement was done using a W tip. ($V_{\text{Set}}=-4$~mV, $I_{\text{Set}}=~200$ pA, $V_{\text{modulation}}=10$ $\mu$V). }
		
		\label{FigS5}
	\end{figure}
	\pagebreak
	\section{Multiple points spectroscopy}
	
	\begin{figure}[!ht]
		\includegraphics[trim= 0cm 3cm 0cm 2cm,clip=true,width=1\textwidth]{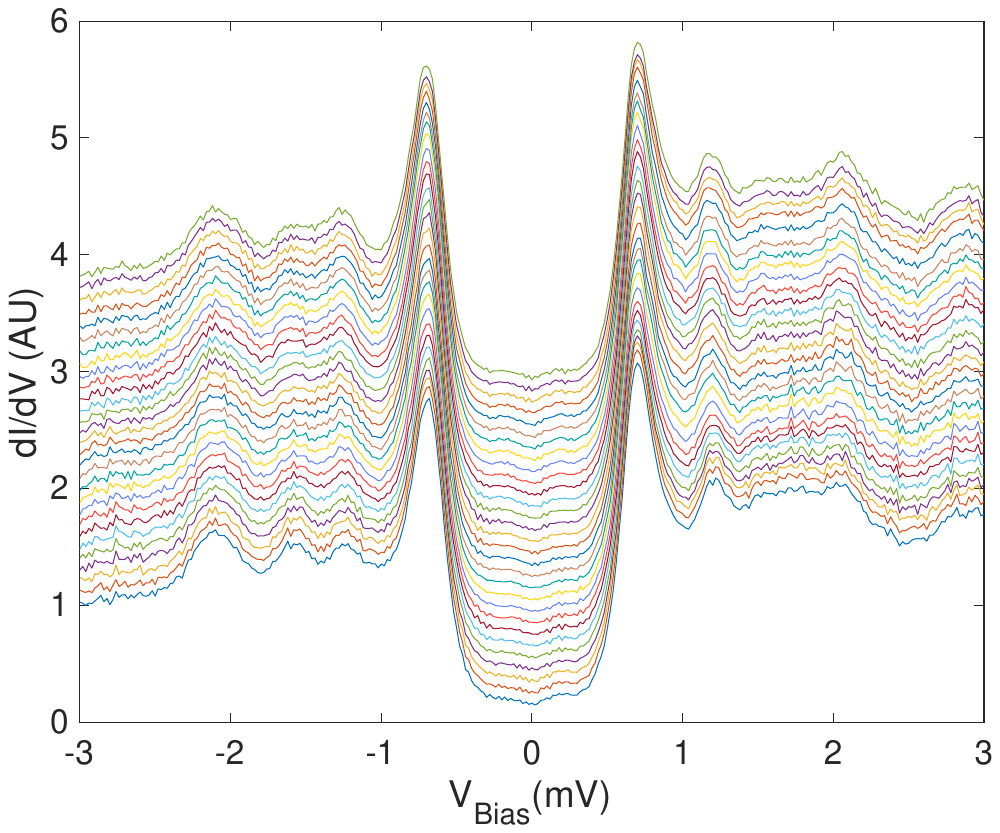}
		\caption{\textbf{Spectra at different point at $x=0.3$.} Superconducting gap spectra extracted form 300 individual points measured on the surface of \RVS\ ($x=0.3$) shown in Fig.~1. The waterfall plot shows that the gap is uniform across the sample and is not influenced by individual Sn dopant sites. It also shows the persistence of the second, smaller gap. ($V_S=-3$~mV, $I_t=150$~pA, $V_{\mathrm{mod}}=30~\mu$V)). }
	\end{figure}

	\pagebreak
	\section{Quasi-Particle Interference data}
	The quasi-particle interference (QPI) pattern in all samples is reminiscent of the data acquired on the parent compound, owing to similar band structure (Supplementary Figure S6). In all doping levels the arc formation (Q2, Q3) lack 6 fold symmetry as was shown before\cite{zhao2021cascade}. The Q1 vector originates from scattering of electrons residing in the Sb band around the $\Gamma$ point, while Q2 and Q3 wave vectors form arcs consistent with those of the V-derived Fermi surfaces reported in CsV$_3$Sb$_5$ at comparable energies \cite{zhao2021cascade}. To test the two-gap scenario in $x=0.3$, we performed QPI measurements inside the superconducting gap. In a multi-band superconductor with distinct gaps, QPI evolve with energy within the superconducting gap~\cite{hoffman2002imaging,wang2003quasiparticle,chen2022evidence,zhang2009quasiparticle}, vanishing at the Fermi level for a fully-gapped order parameter. For \RVS\, such data is presented in Supplementary Figure S6. 
	
	For energies larger than the large gap ($\Delta_1$), the QPI reproduces the normal-state band structure similar to other \RVS\ doping levels, and evolves with energy within the superconducting gap ($E<\Delta_1$). For energies similar to the major coherence peaks energy, scattering is dominated by the signal from the centered ellipse, the Sb derived band. The QPI signal centered around the $\Gamma$ point has a $q$ vector of $2.8~\text{nm}^{-1}\pm0.1$, which matches electron scattering in the Sb band around the $\Gamma$ point, as seen by ARPES and other STM studies~\cite{kato2023surface,li2022rotation,liang2021three,hu2023electronic}. At energies near the energy of the minor coherence peaks, arcs at larger $q$ become prominent, and the ellipse around the center completely disappears. Finally, at the Fermi level, the QPI signal vanishes, consistent with a fully gapped order parameter. This systematic evolution provides independent spectroscopic evidence for two distinct superconducting gaps. Furthermore, this implies that the larger gap resides on the planar Sb-dominated pocket around the Gamma point, whereas the smaller gap resides on the large Fermi surface due to the V atoms.
	
	\begin{figure}[!ht]
		\includegraphics[trim= 0cm 0cm 0cm 0cm,clip=true,width=1\textwidth]{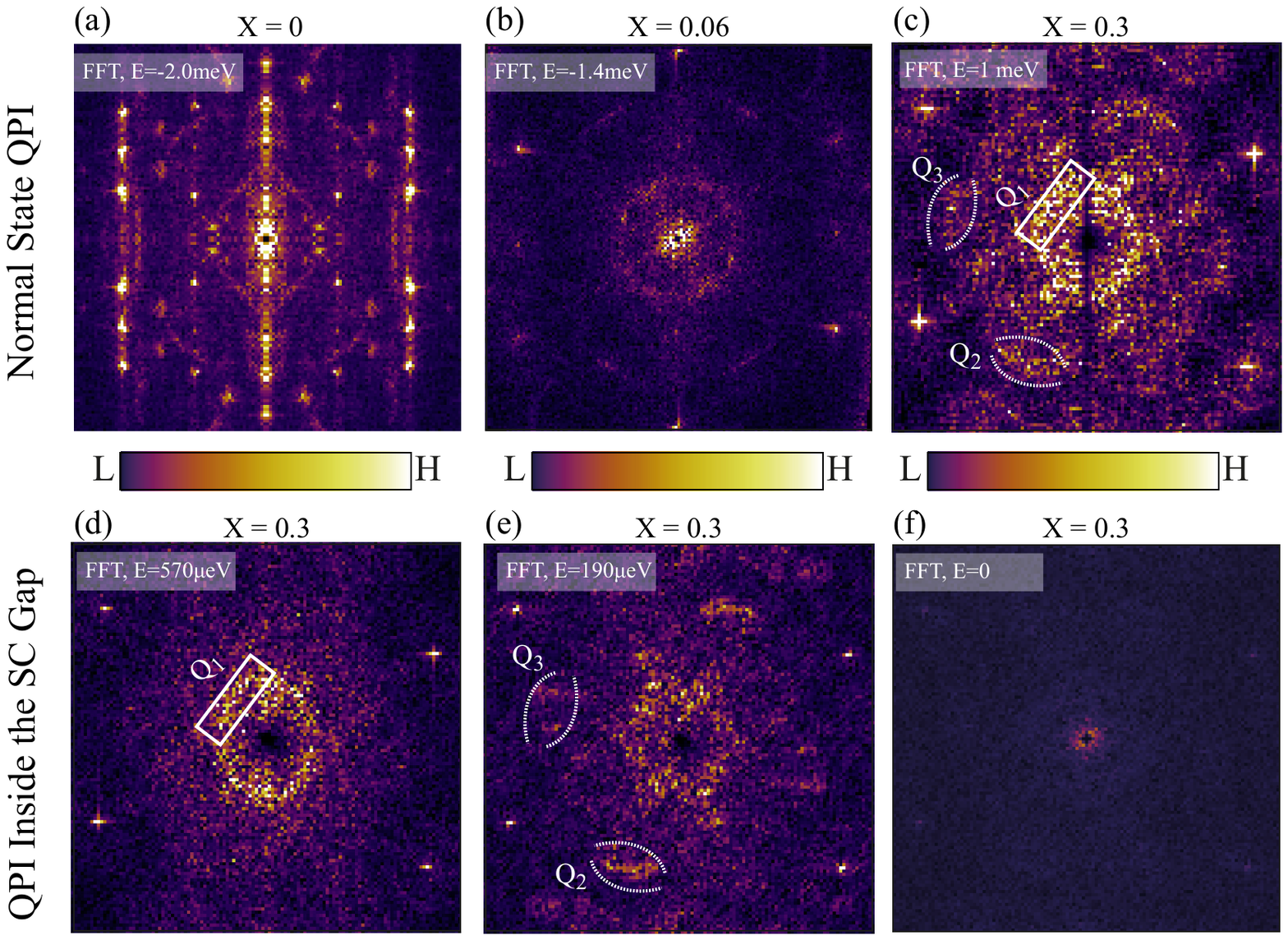}
		\caption{\textbf{QPI data for different doping levels.}  \textbf{a} QPI data acquired on the surface shown in Supplementary Figure 4. The data is symmetrized along the Y-axis of the FFT figure, and is reminiscent of the data given in other studies such as CsV$_3$Sb$_5$~\cite{chen2021roton,zhao2021cascade}. The data is taken at 300~mK, and shows scattering of quasi particles in the superconducting state, at energy outside the superconducting gap. \textbf{b} QPI data acquired on the surface shown in Figure 1 ($x=0.06$). The data is non-symmetrized, and show similar features to the parent compound, such as the main scattering q vectors around the center and arcs between the Bragg peaks. The data is taken at 300~mK, at energy outside the superconducting gap. \textbf{c} QPI data acquired on the surface shown in Figure 1 ($x=0.3$), with Q1 gives a scattering vector of $2.8~\text{nm}^{-1}\pm0.1$ and the expected scattering vector for the Sb band given from a linear interpolation of the DFT data gives $2.6~\text{nm}^{-1}$. The data is non-symmetrized, and show overall similar features to the parent compound and lightly doped samples. such as the main scattering q vectors around the center and arcs between the Bragg peaks. The data is taken at 300~mK, at energy outside the superconducting gap. \textbf{d-f} QPI data acquired on the surface shown in Figure 1 ($x=0.3$). The data is non-symmetrized, and show the evolution of the scattering vectors, shown in \textbf{c}, with energy. The data was acquired at energies smaller than the superconducting gap. The scattering vectors are seen next to the large coherence peaks (d), then only the arcs at larger q vectors remain (e), and finally all scattering is absent at the Fermi level (f). These shows points on the portions of the Fermi surface associated with the two gaps. Vanishing of the signal at the Fermi level points to a fully gapped superconductor. For a-b ($V_{\text{Set}}=4$~mV, $I_{\text{Set}}=~200$ pA, $V_{\text{modulation}}=150$ $\mu$V), For c-f ($V_{\text{Set}}=-3$~mV, $I_{\text{Set}}=~150$ pA, $V_{\text{modulation}}=200$ $\mu$V).}
		
		\label{FigS6}
	\end{figure}

	
	\clearpage
	\section{Superconducting conductance maps around the defect at the Fermi Level}
	For the impurity shown in Figure 2 of the main text, we show here a $dI/dV$ map of an $E=0$ meV energy slice, acquired on and around a type I defect. Even though $dI/dV$ the map is not exactly at the bound state energy, it shows a strong signature of the bound states.
	
	\begin{figure}[!ht]
		\includegraphics[trim= 0cm 0cm 0cm 0cm,clip=true,width=0.75\textwidth]{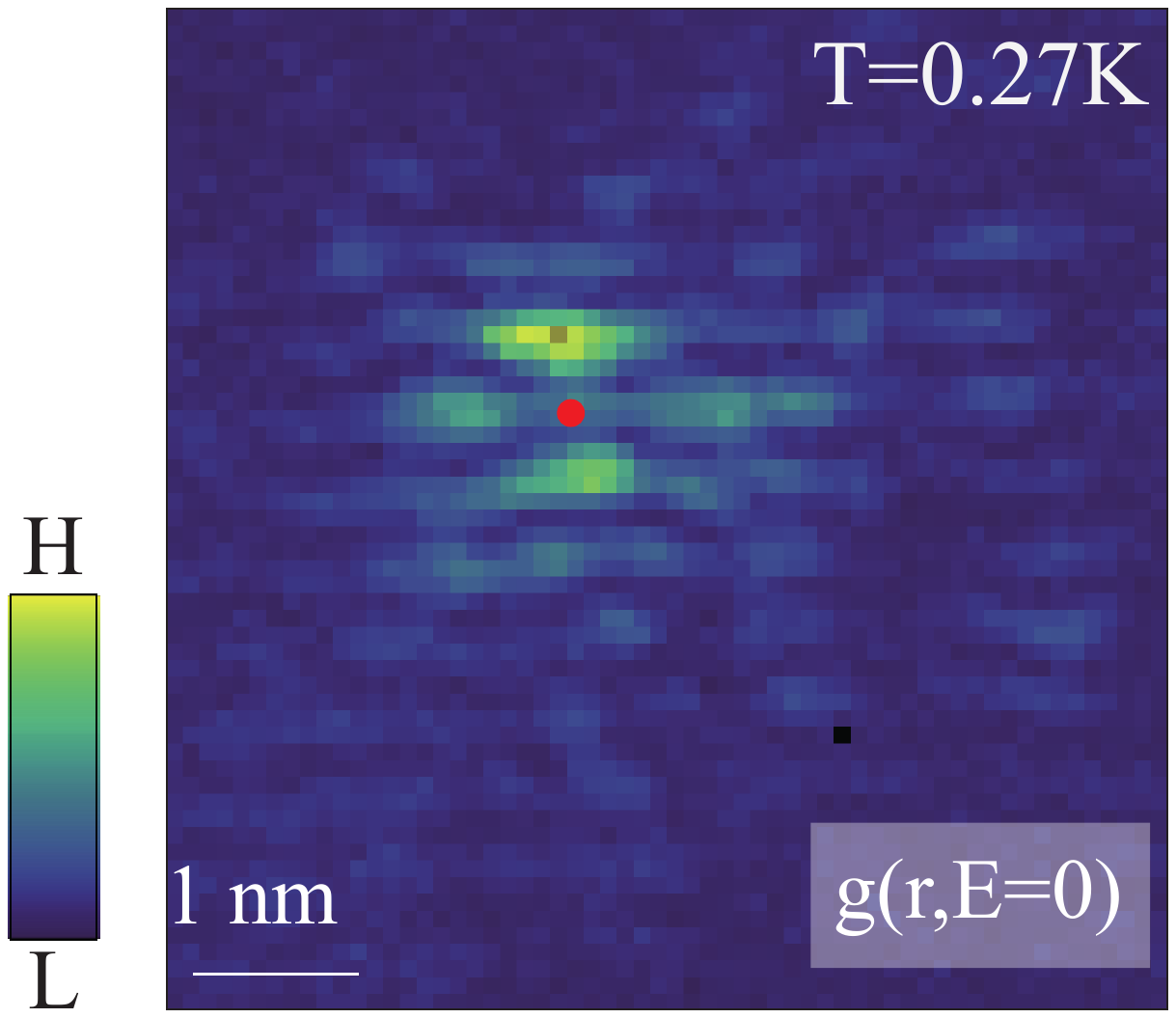}
		\caption{\textbf{Conductance map at the Fermi level around the Type I defect} Conductance map around the defect shown in Figure 2 of the main text, a type I defect. The signature of the bound state is seen easily in the energy slice at the Fermi level, as opposed to maps of lower doping levels. ($V_S = -3$~mV, $I_t = 150$~pA, $V_{\text{mod}} = 30$ $\mu$V). The temperature is $T=0.27$ K.}
		\label{FigS7}
	\end{figure}
	
	\clearpage
	\section{Systematic analysis of the YSR states for charge and magnetic impurity scattering}
	
	\begin{figure}[!ht]
		\centering
		\includegraphics[width=\linewidth]{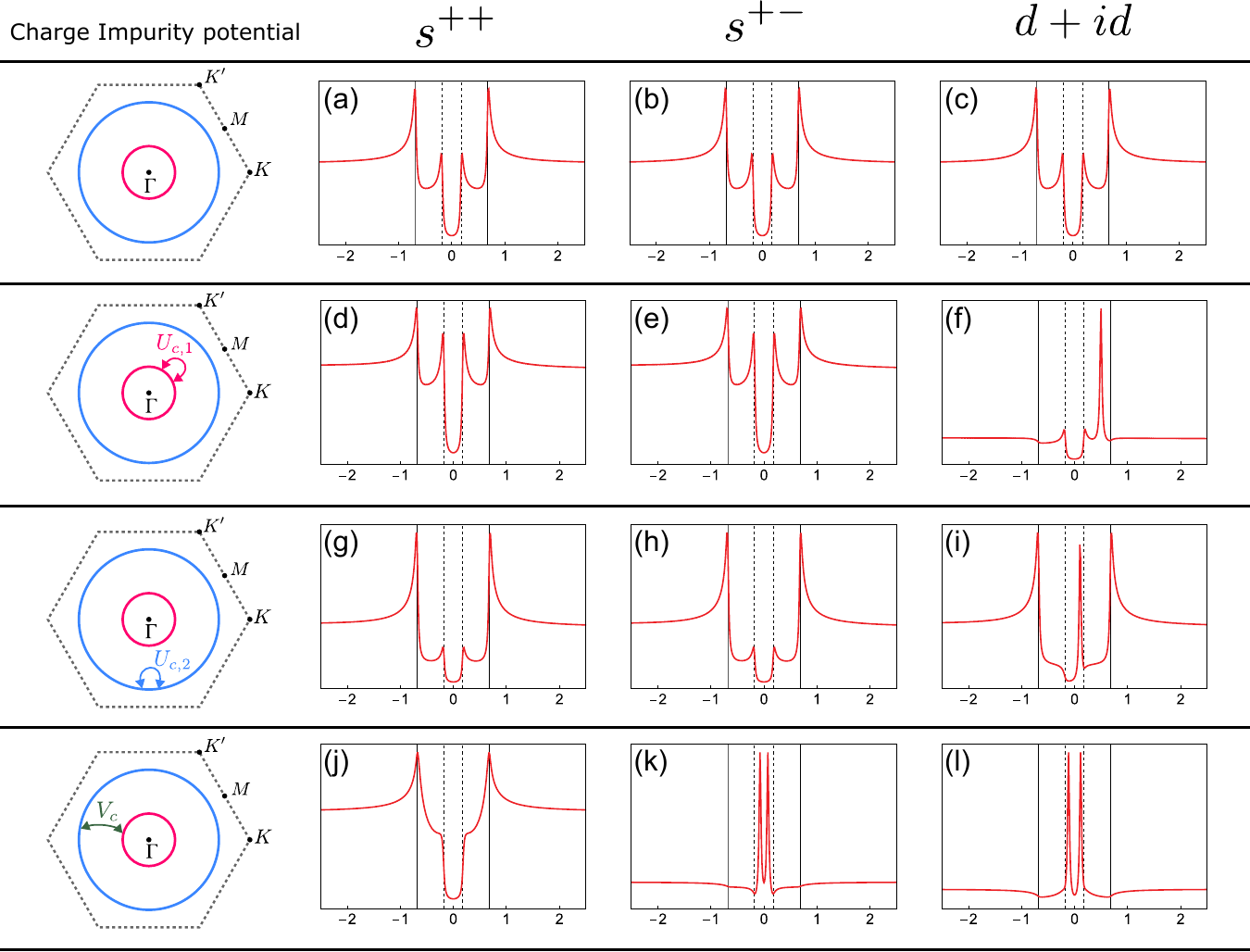}
		\caption{\textbf{Calculations for a nonmagnetic (charge) impurity embedded in superconductors with different symmetries.} Local density of states at the impurity position for $s^{++}$, $s^{+-}$ and $d+id$ superconducting gaps and for a nonmagnetic (charge) impurity potential. (a)-(c): without impurity, (d)-(f): with impurity potential ($U_{c,1}$) within Fermi surface 1 (FS1), (g)-(i): with impurity potential ($U_{c,2}$) within Fermi surface 2 (FS2), (j)-(l): with impurity potential between two Fermi surfaces ($V_c$).}
		\label{fig:placeholder}
	\end{figure}
	
	\begin{figure}[!ht]
		\centering
		\includegraphics[width=\linewidth]{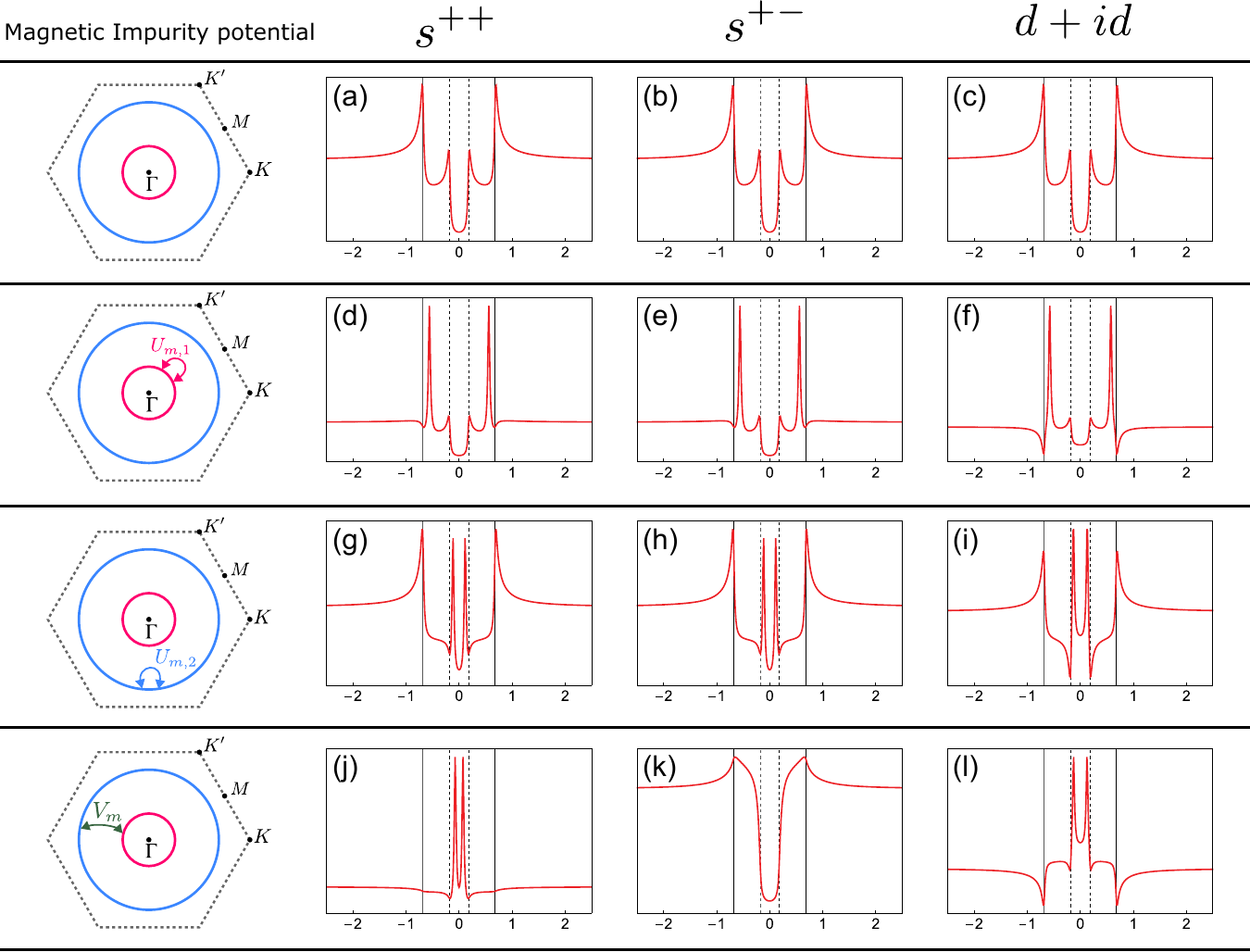}
		\caption{\textbf{Calculations for a magnetic impurity embedded in superconductors with different symmetries.} Local density of states at the impurity position for $s^{++}$, $s^{+-}$ and $d+id$ superconducting gaps and for amagnetic impurity potential. (a)-(c): without impurity, (d)-(f): with impurity potential ($U_{m,1}$) within Fermi surface 1 (FS1), (g)-(i): with impurity potential ($U_{m,2}$) within Fermi surface 2 (FS2), (j)-(l): with impurity potential between two Fermi surfaces ($V_m$).}
		\label{fig:placeholder}
	\end{figure}
	
	\clearpage
	
	\section{$dI/dV$ maps and spectra of additional defect sites}
	
	\begin{figure}[!ht]
		\includegraphics[trim= 0cm 0cm 0cm 0cm,clip=true,width=0.8\textwidth]{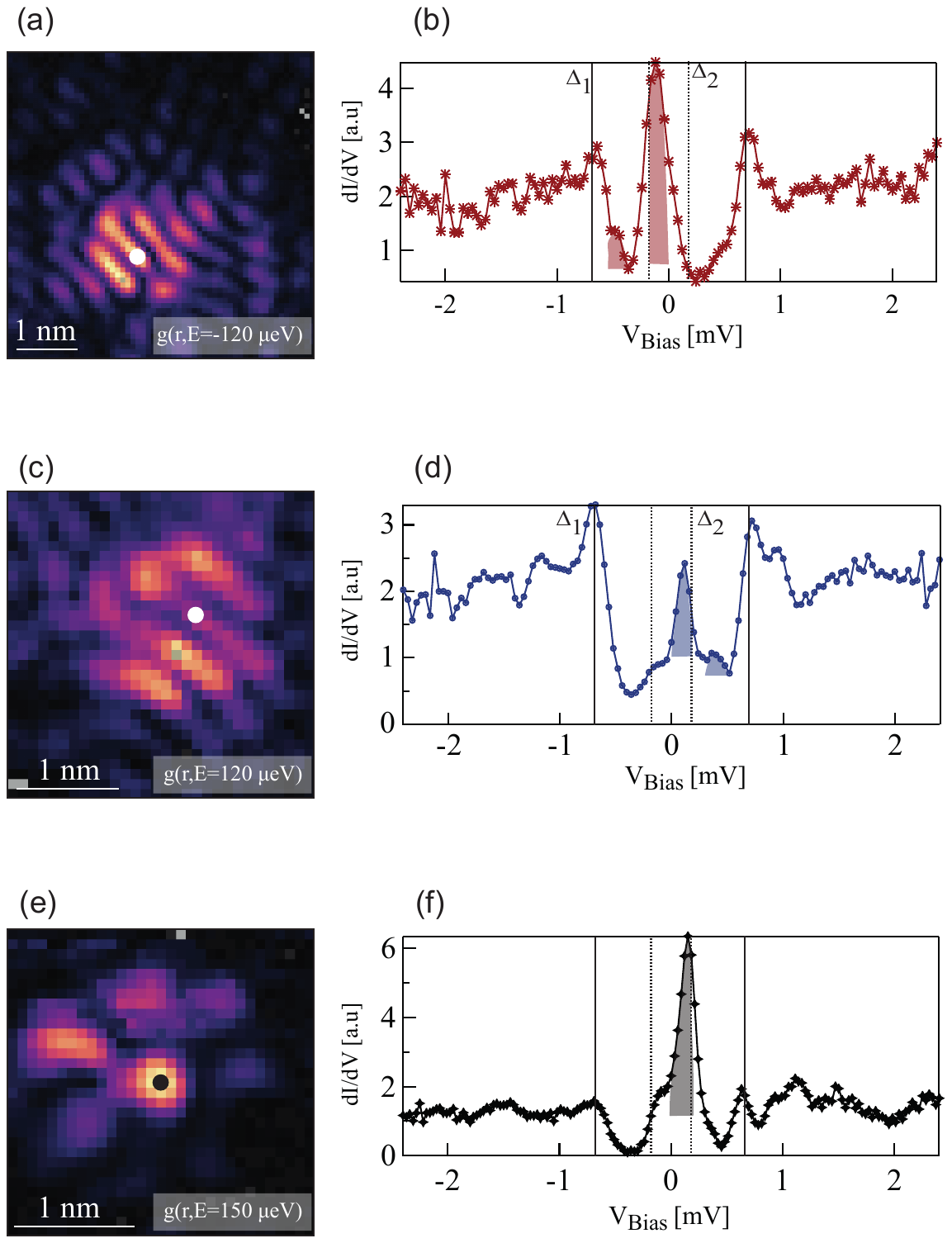}
		\caption{\textbf{Additional defect sites.} \textbf{a,b} Additional defect site measured in \RVS. A conductance map (a) and a spectrum (b) are shown around and at the center of the defect respectively. This defect is of type I, where a peak is seen in the bare spectrum at $-120\,\mu$eV, and $-480\,\mu$eV, and a clear $C_2$ symmetry of the oscillations. \textbf{c,d} Same as a,b with energies $120\,\mu$eV, and $400\,\mu$eV. \textbf{e,f} A defect of type II, with a single peak present at $150\,\mu$eV. This defect lacks a bound state at energies above $\Delta_2$, and the $C_2$ symmetry of the bound state with a more circular halo around the center. Parameters for a,b: $V_S=-4$ mV, $I_t=200$ pA. c,d: $V_S=-4$ mV, $I_t=200$ pA. e,f: $V_S=-3$ mV, $I_t=150$ pA}
	\end{figure}

	\clearpage
	\section{Relationship between normal state DOS asymmetry and YSR state asymmetry: $s$-wave with magnetic impurity}
	
	Here we consider the effect of an asymmetric normal states density of states on the asymmetric peak height of the YSR states in an $s$-wave superconductor with a magnetic impurity. A possible origin of asymmetric YSR spectra is a strongly asymmetric normal-state density of states. To examine this possibility, we modeled a conventional $s$-wave superconductor containing a magnetic impurity embedded in a normal state with an asymmetric density of states (Supplementary Fig.~12). These calculations show that reproducing the experimentally observed asymmetry requires an unrealistically strong asymmetry already in the normal state. Experimentally, however, the tunneling spectra measured above both impurity types exhibit nearly featureless normal-state spectra with no comparable asymmetry. We therefore conclude that the observed particle-hole asymmetry cannot originate from the underlying normal-state electronic structure.
	
	To model an asymmetric density of states of the normal state, we use a following formula:
	\begin{align}
		\rho(\epsilon)=\rho_0\left[1+\lambda\tanh\left(\frac{\epsilon}{\Lambda}\right)\right]
	\end{align}
	where we set $\rho_0=1$. 
	
	\begin{figure}[!ht]
		\centering
		\includegraphics[width=\linewidth]{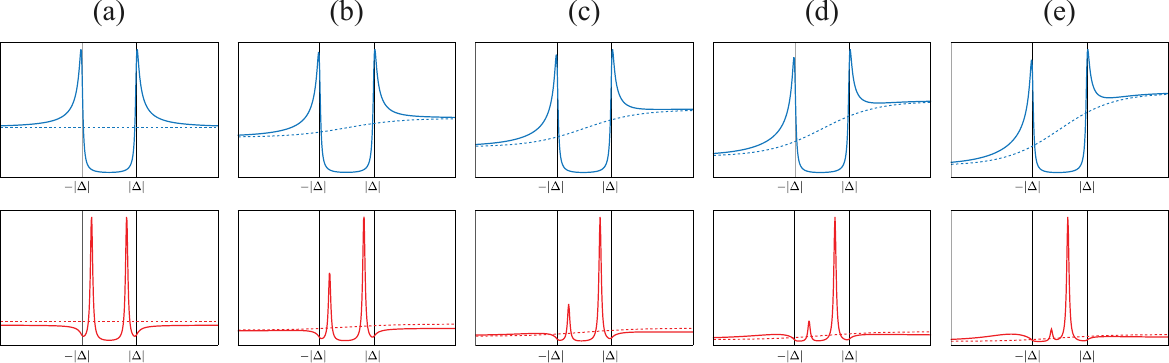}
		\caption{\textbf{Calculations for an asymmetrical DOS in the normal state, with a magnetic impurity embedded in an s-wave superconductor.} The local density of states for the normal state are shown by dashed lines, whereas for the superconducting state are shown by the dashed lines. Blue corresponds to the case without impurity and red, to the case with a magnetic impurity. The asymmetry parameters used are: (a) $\lambda=0$, (b) $\lambda=0.2$, (c) $\lambda=0.4$, (d) $\lambda=0.6$, (e) $\lambda=0.8$ with $\Lambda=2|\Delta|$.}
		\label{fig:placeholder}
	\end{figure}

	
	
	


\end{document}